# Guide to *k*-mer approaches for genomics across the tree of life

Katharine M. Jenike[1], Lucía Campos-Domínguez[2], Marilou Boddé[3], José Cerca[4], Christina N. Hodson[5], Michael C. Schatz[1], Kamil S. Jaron[3]
[1] Johns Hopkins University, School of Medicine, Baltimore MD USA
[2] Centre for Research in Agricultural Genomics, CRAG (CSIC-IRTA-UAB-UB), Campus UAB, Cerdanyola del Vallès, 08193 Barcelona, Spain
[3] Tree of Life, Wellcome Sanger Institute, Cambridge CB10 1SA, UK
[4] Centre for Ecological and Evolutionary Synthesis, Department of Biosciences, University of Oslo, Norway
[5] Department of Zoology, Biodiversity Research Centre, University of British Columbia, Vancouver, BC, V6T 1Z4, Canada

**Abstract**

The wide array of currently available genomes display a wonderful diversity in size, composition and structure with many more to come thanks to several global biodiversity genomics initiatives starting in recent years. However, sequencing of genomes, even with all the recent advances, can still be challenging for both technical (e.g. small physical size, contaminated samples, or access to appropriate sequencing platforms) and biological reasons (e.g. germline restricted DNA, variable ploidy levels, sex chromosomes, or very large genomes). In recent years, *k*-mer-based techniques have become popular to overcome some of these challenges. They are based on the simple process of dividing the analysed sequences (e.g. raw reads or genomes) into a set of sub-sequences of length *k*, called *k*-mers. Despite this apparent simplicity, *k*-mer-based analysis allows for a rapid and intuitive assessment of complex sequencing datasets. Here, we provide the first comprehensive review to the theoretical properties and practical applications of *k*-mers in biodiversity genomics, serving as a reference manual for this powerful approach.

# Introduction

The genomics field has come a long way in the past quarter century. Sequencing and assembling even a partial genome was once a multi-billion dollar accomplishment [1]. Now complete telomere-to-telomere (T2T) assemblies are being produced for an ever increasing array of species [2] [3] [4] and large global efforts to capture the genomic diversity of life are well underway [5]. However, as of this writing the vast majority of known species do not have an assembly, or even sequencing data associated with them. In fact, only 1% of known species have an associated assembly according to the Genomes on a Tree tracker [6]. Genome assembly is however a very complex process that needs to be guided by orthogonal assembly-free approaches. These methods are frequently based on a popular bioinformatic concept "*k*-mers". *K*-mers have proven to be an efficient and powerful concept for understanding the vast and continually growing sequencing data.

Every genomicist has encountered *k*-mers in one way or another. While the names have changed substantially over the years (See **Box 1**), the concept remains the same. For example, *k*-mers are used in the back-end of many bioinformatic alignment tools such as BLAST [7], bowtie2 [8], and minimap2 [9] and practically all genome assemblers (See [10] for a review). More recently, *k*-mers have emerged as a popular framework for analysing genomic datasets directly in methods such as genome profiling, thanks to the development of user-friendly tools such as GenomeScope [11] [12]. Consequently, *k*-mers are now a fundamental part of genomics, which motivates this guide. We define and explain the basic properties of *k*-mers, review the established *k*-mer-based techniques in genomics, and showcase recent creative uses of *k*-mers for complex genomic problems. Together with the manuscript, we provide online materials for the reader to exercise the knowledge on real biological examples available here: https://github.com/KamilSJaron/oh-know/wiki

## Box 1: Historical perspective on *k*-mers

The oldest reference to the concept dates back to the legendary work of Claude Shannon [13], where he used "**N-grams**" to develop a theory for communication, later to calculate entropy of a natural language. In the mathematically oriented research community, the concept is most often referred to as "***k*-tuples"** [14] [15], including shorter versions such as "**ktup"** [16], or other prefixes "**L-tuple**" [15] or "**ℓ-tuple**" [17]. Perhaps to reach a wider audience, some authors decided to use "***k*-word**" [17,18] or just "**word**" [19]. Instead, it was "***k*-mer**" that became the most popular and common expression. Sequencing by hybridization used "**11-mers**" for the oligo templates [20], although they never used the generalised form with *k*. Although the concept of *k*-mers also appeared in the publication of BLAST as "**w-mers**" [7,20], it took nearly a decade before *k*-mer became more commonly used. The use of "*k*-mer" became more common in late 1990s, including within the seminal work of the whole genome aligner MUMmer in 1999 [21]. In 2000, Liu & Singh coined "*k*-mer word frequency distribution" and described it as a "signature" of the sequence [21,22]. A few years later, Mullikin & Ning published a "word frequency graph", which is the first record of a *k*-mer spectrum [23]. Publications using the word *k*-mer increased in the following years compared to any other of the terms. The expression "*k*-mer" was solidified as the main way to describe this concept throughout the 2000s with the release of several genome assemblers, read aligners, and specialised software for counting kmers such as Jellyfish [23,24]. See **Supplementary Text 1** for a few more details.

## *k*-mer basics

In a genomic context, *k*-mers are sub-strings of nucleotides of length *k* contained within a sequence (e.g. individual reads, reference genome, or any other sequence). *K*-mers are typically used for DNA but the concept can be applied to RNA or protein sequences as well. Any genomic sequence can be decomposed into a number of consecutive *k*-mers, and this number will depend on both the length of the sequence (L) and *k*-mer length (*k*). For example, in the following sequence: AAGTCCAT (L=8), there are seven *k*-mers of length 2 (2-mers), six 3-mers, five 4-mers, four 5-mers, three 6-mers and two 7-mers (**Figure 1**). The number of *k*-mers in a sequence of length L is equal to  L - *k* + 1. This is a general principle that can be applied to any sequence, regardless of the sequence length or composition.

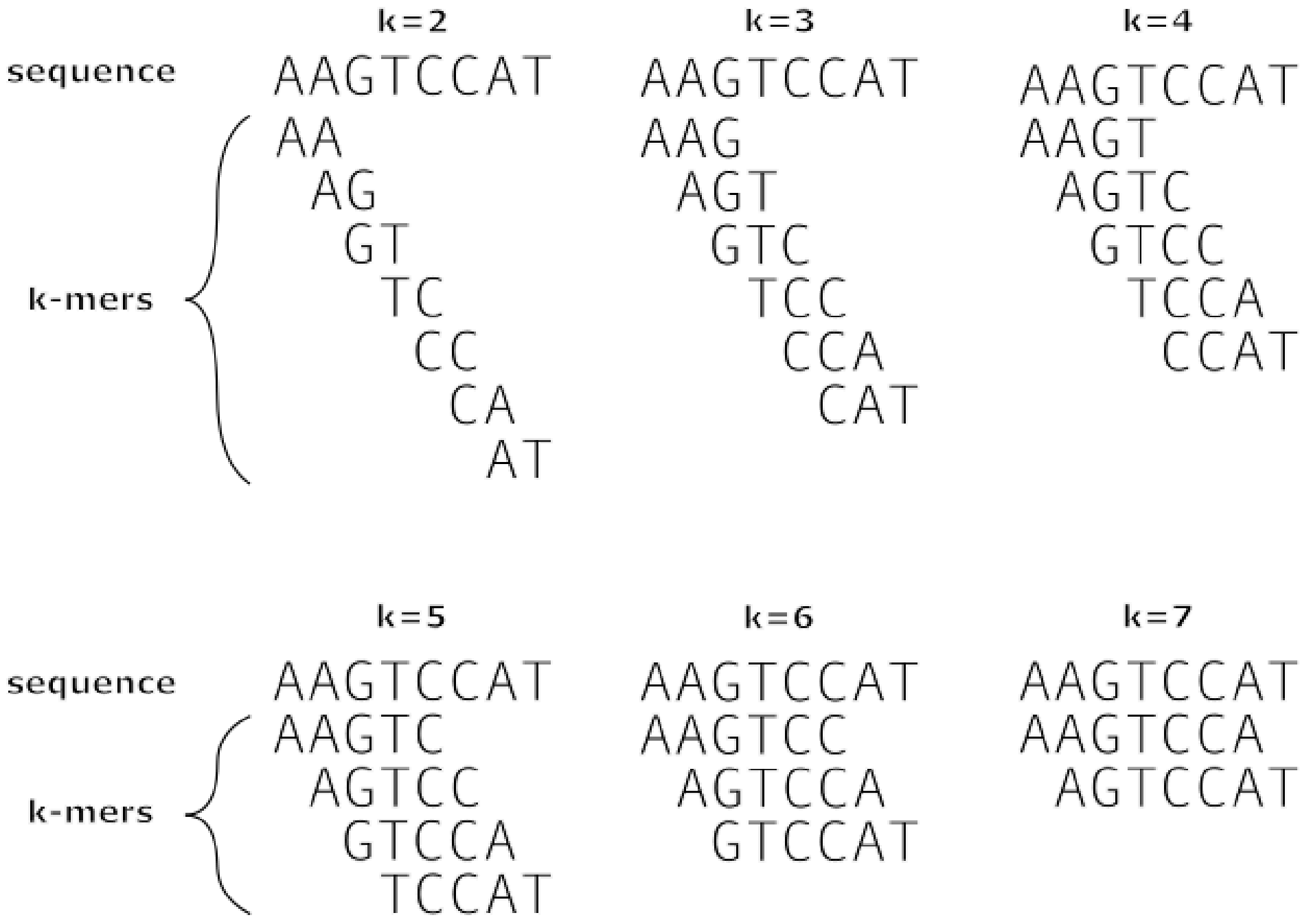

**Figure 1**: **Simple example of *k*-mer decomposition.** The example sequence decomposed into *k*-mers for a range of *k* values, from 2 to 7. Notice the number of *k*-mers in a read is L - *k* + 1.

What is the point of further fragmentation of sequencing reads or genomes? At the most fundamental level, they provide a convenient way to break apart a genomic sequence into simpler words. Unlike natural language, which has the benefit of spaces and other punctuation marks, genomes don't intrinsically mark the start or end of words. Nevertheless *k*-mers impose such a structure with surprisingly powerful outcomes. One major benefit is improving the signal-to-noise ratio. One sequencing error in a read means that the read is not a perfect match to the sequenced template, but only a subset of its *k*-mers will inherit the error. A read of length L contains L - *k* + 1 *k*-mers, but only *k* of them will contain the incorrect sequence (e.g. for 100bp reads analysed with 21-mers, a sequencing error will disrupt 21 of the 80 *k*-mers in the read). The remaining *k*-mers will represent true genomic sequence. The second benefit is purely computational - operating on the premise we would like to analyse the *k*-mers that are correct, we can use exact matches to count *k*-mers, which is much faster than any imperfect matching algorithm (e.g. using a hash table, or binary search instead of computing a complete alignment). Most notably, *k-mers* enable genome assembly via de Bruijn graphs used in assemblers such as EULER [25] or spades [26]. De Bruijn graphs are constructed of *k*-mers found in the reads, followed by a series of sophisticated graph transformations [27]. Finally, a third benefit is that *k*-mers can be used to rapidly assess various biological features. For example, they can be used to estimate genomic properties such as genome size, genome repetitiveness, and heterozygosity [11]. They can also be used to estimate similarity between genomes without alignment [28]. Before expanding further, however, we need to understand the properties of *k*-mers and how the choice of *k* affects the dataset.

## Essential properties of *k*-mers

The choice of *k* affects two essential properties of *k*-mers: the complexity of the *k*-mer set and the *k*-mer coverage. Here, the complexity is the number of possible distinct *k*-mers given *k*. With a four base pair alphabet, this complexity is $4^k$. However, in practice, we usually disregard the sequence strand and, as a result, the reverse complement sequences are typically counted as the same sequence as the forward sequence, e.g. CAT and ATG would be counted together. These collapsed *k*-mer pairs are called "canonical *k*-mers", and are typically implemented by selecting the lexicographically smaller of the *k*-mer and its reverse complement (e.g. ATG would also represent CAT) as the sequence for the pair. Consequently, for an odd value of *k*, there are $4^k$ / 2 possible canonical *k*-mers (See [29] for more details). In the remainder of this paper, unless stated otherwise, *k*-mers refer to canonical *k*-mers.

The complexity of the *k*-mer space (all possible *k*-mers) increases exponentially with *k* (**Supplementary Figure 1**). However for low values of *k*, this space will be largely insufficient to represent or distinguish long sequences. For *k = 3*, there are only 32 possible *k*-mers given a 4 base pair alphabet, for *k* = 7, there are 8,192; for these low *k* values all possible *k*-mers will appear in a genome many times, and all these *k*-mers will be most likely observed in any species. However, the relative frequencies of these short *k*-mers ($k \leq 11$) , sometimes called genomic signature, still carries a phylogenetic signal ([30] [31], see [32] for a review).

For most applications we need to select *k* long enough so that most *k*-mers in the genome will not be repetitive. The proportion of *k*-mers corresponding to a unique position in the genome increases with greater *k* (see **Figure 1** of [33]). For humans, the smallest *k* with some *k*-mers unique to a specific genomic location is 11, although such space contains very few unique positions (less than $2.1 \times 10^6$ *k*-mers out of $3 \times 10^8$ possible positions [34]). Instead, we estimate the minimum length of *k* by computing the expected number of occurrences of a given *k*-mer in a genome of length G. This is estimated as $(G/4)^k$, meaning we need to select *k* to be at least $\log_4(G)$. For the 3Gb human genome, this would suggest a minimum length of 16; practically however, the human genome has many 16bp repeats so it is advantageous to select even longer *k*-mers.

With *k* large enough to contain a substantial amount of *k*-mers that are unique to a single position in a genome, we can estimate how many reads on average contain a *k*-mer found in a single copy in the genome, which is the definition of the mean *k*-mer coverage $C_k$. The classical definition of sequencing coverage $C_g$ is the mean read depth per base in a genome

$$C_g = N * L / G,$$

where N is the number of sequenced reads, L is the average read length and G is the total genome size. The average genome coverage and average *k*-mer coverage are not the same, however, there is a simple relationship approximating their relative values:

$$C_k = C_g(L - k + 1) / L$$

So for a given read set, lower values of *k* give a higher *k*-mer coverage, and the differences in *k*-mer coverage for different values of *k* will be more apparent for short reads (**Supplementary Figure 2**). For example, using *k*=100 for 100bp reads is problematic as a read is represented by a single *k*-mer but using *k*=51 allows for 50 *k*-mers per read.

Finally, the process of sequencing has limited accuracy that varies by sequencing platform, and a certain fraction of the sequenced nucleotides will represent sequencing errors. These errors will propagate from reads to *k*-mers, and the fraction of affected *k*-mers is again dependent on the chosen value of *k.* Assuming uniformly distributed errors (i.e. each sequenced position within each read has the same probability of being wrong), the probability of a *k*-mer representing the real genomic sequence is then

$$\Pr(\text{error free } k\text{-mer}) = (1 - e)^k$$

where *e* is the sequencing error rate per base. Therefore, the lower *k* we chose, the greater the proportion of *k*-mers that will represent the true genome sequence (**Supplementary Figure 2**) and therefore be useful for downstream analyses.

Altogether, the choice of *k* is a trade-off between captured genomic complexity, sequencing coverage, and error rates. Therefore the ‘right’ value of *k* depends on the application, sequencing depth, sequencing platform and type of genome sequenced. For example, long read sequencing technologies allow using higher values of *k*, because the *k*-mer coverage will not decrease dramatically for higher values of *k* (see the equation above) allowing for $k >$ 1000 for certain applications. This being said, the sequencing error rates still present a limitation on the practical choice of *k*. For short read datasets, a *k* in the range of 21 to 31 bases is typically chosen as it generates large numbers of *k*-mers from unique positions of a sequenced genome for almost any organism, while still providing high *k*-mer coverages. Furthermore, using $k \leq 31$ is more computationally efficient than longer values of *k*, since these sequences can be represented in less than 64 bits, making it fast for computers to compare them (i.e. instead of comparing them character by character, they can be compared in a single computer instruction by treating them as 64-bit integers). Finally, for some analyses it is also beneficial to select odd values for *k*, so that forward and reverse *k*-mers will have distinct sequences, e.g. the reverse complement of AT is also AT, but the reverse complement of ACT will be AGT.

## Analysis of sequencing libraries using *k*-mer spectra

One of the most common direct applications of *k*-mers in genomics is for characterization of a sequencing dataset using the distribution of *k*-mer coverages; this is commonly referred to as *k*-mer spectrum or *k*-mer histogram. A typical *k*-mer spectrum of a moderately heterozygous diploid organism features four apparent coverage peaks (**Figure 2**): The first peak represents sequencing errors (those with low coverage; in pink on the figure); the second peak represents unique genomic sequences from heterozygous loci (1n; yellow). This peak will be centred around $C_k$ coverage, sometimes referred as 1n or monoploid *k*-mer coverage; The third peak represents all homozygous loci in the genome centred around $2*C_k$ coverage (2n; blue); and the fourth, usually a much smaller peak represents genomic duplications centred around $4*C_k$ coverage (4n; orange).

The *k*-mer spectra analysis has become a standard step of genome sequencing [35] and the need for efficient tools was recognised by the bioinformatics community. As a result, there has been a significant improvement in performance of *k*-mer counting tools with many additional functionalities developed in *k*-mer toolkits [24] [36] [37] [38] [39] [40] (See an overview of popular *k*-mer counters in the **Supplementary Table 1**).

Calculating the *k*-mer spectrum is however just the first step. Various models can be fitted to the *k*-mer spectrum to estimate genomic features such as genome size, heterozygosity or repetitiveness, collectively called genome modelling or genome profiling.

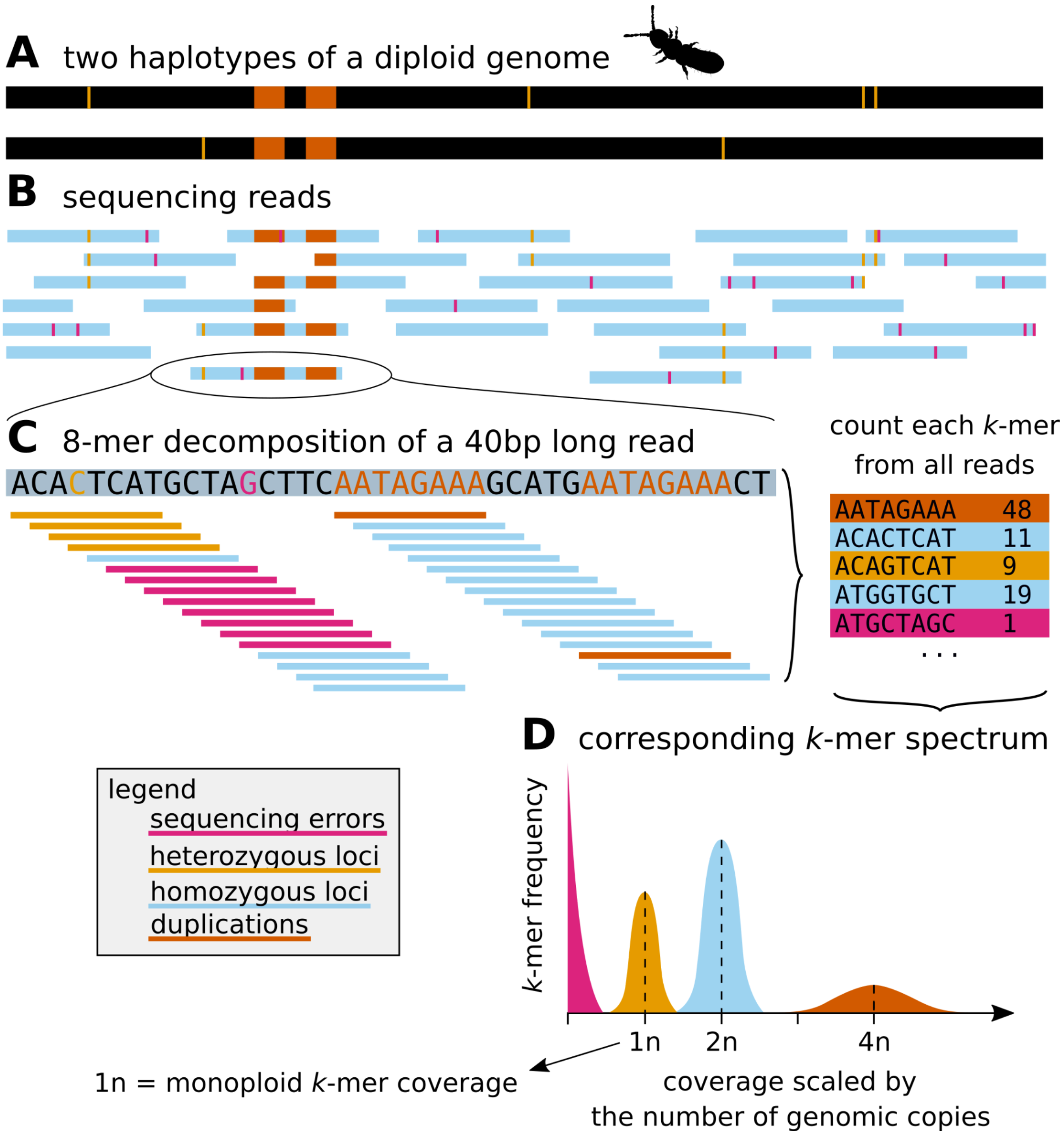

**Figure 2: From sequencing reads to *k*-mer spectrum.** This figure shows a basic illustrative example of how genomic reads can be translated into a number of *k*-mers that can be counted and represented by a *k*-mer spectrum. **A)** Genome representation in two haplotypes with a duplication (orange) and six heterozygous loci (yellow). **B)** The sequencing reads and their alignment to the genome contain the genomic sequence but are also overlaid with sequencing errors (pink). **C)** Display of an *k*-mer (8-mer) decomposition for a 40 bases long sequencing read **D)** *k*-mer spectra represent how many different *k*-mers (y-axis) show a specific coverage (x-axis) in a sequenced genome. In the vast majority of real genomic dataset, the peaks in the *k*-mer spectra would overlap, and furthermore, there would be a small number of *k*-mers representing other ploidies too (e.g. 3n for heterozygous duplications).

## Fitting a genome profiling model

One of the first genome profiling techniques was designed to estimate *k*-mer coverage and genome size [17]. The idea behind genome size estimation is practically the same as calculating *k*-mer coverage (Equations 1 and 2). That is because genome size (G) is the total number of *k*-mers divided by the *k*-mer coverage ($C_k$) and the ploidy level (p).

$$G = N (L - k + 1) / (C_k * p)$$

where N is the total number of reads in the dataset and L is the read length, which together with *k-mer* size *k* give us the total number of *k*-mers in the dataset. This equation ignores sequencing errors, which will artificially inflate the estimated genome size by counting *k*-mers introduced from sequencing errors as part of the genome. Tools that include an error model can subtract the estimated error *k*-mers, which allows for a more precise genome estimate. The estimation of the *k*-mer coverage ($C_k$) is done by fitting one, or multiple, distributions to the empirically calculated *k*-mer spectra. The simplest distribution, and a natural choice for a sampling process such as genome sequencing, is the Poisson distribution [17,41,42]. A Poisson distribution is easy to fit because it is characterised by a single parameter, determining both the mean and the variance. However, in practice the observed variance is usually greater than the variance fitted by the Poisson distribution (i.e. the observed distribution is 'overdispersed') so that it might be more appropriate to fit a negative binomial distribution [11,12,43]. The latter is a generalisation of the Poisson distribution and is characterised by two parameters, a mean parameter and a separate variance parameter, but has a similar shape. A normal or Gaussian distribution can be used for high coverage values, but is not appropriate for low coverage (below 10x coverage). This is because a Gaussian distribution is symmetric around the mean, so that at low coverage it will estimate coverage values to be less than 0 (negative values) that are not sensical. An alternative distribution, that particularly addresses the fact that some sites have a lower probability of being sequenced than others (e.g. heterochromatin), is the skew normal distribution. Authors of findGSE proposed skew normal distribution to address this [44]. No distribution fits all purposes, but in general, the more complicated coverage models are better suited for high coverage datasets, while in the case of low coverage data it is preferable to use the least parameterized distribution [41].

The majority of model-fitting approaches expect relatively high sequencing coverage (>10x) so that they can fit the parameters of the distributions with very few assumptions. Authors of RESPECT took a different route, designed for low-coverage (skimming) data, and specifically optimised for datasets in the range 0.5 - 2x sequencing coverage [41]. The model estimates the number of *k*-mers that are found in one, two, three, etc counts in the genome, while using simple error and coverage models [41]. To make the model fit possible, the authors constrained some of the parameters; for example, errors are modelled as a proportion of *k*-mers observed only once in reads, but also the individual counts of *k*-mers in the genomes are constrained by empirically observed ratios in several hundred publicly available genomes [41]. These empirical constraints were generated using haploid representation of diploid genomes, therefore we would expect reliable estimates only for haploid species or diploid genomes with low heterozygosity.

Interestingly, the *k*-mer spectrum can also reveal other properties of the genome. The genome-wide level of heterozygosity can be determined through an analysis of the number of *k*-mers in the 1n vs 2n peaks: at low rates of heterozygosity, the 1n peak will be relatively small since most *k*-mers will be homozygous, but the 1n peak grows taller with higher numbers of heterozygous *k*-mers. Only a relatively modest rate of heterozygosity is needed to elevate the 1n peak to match the 2n peak since each heterozygous variant will cause 2**k* heterozygous kmers (assuming the variants are spaced out appropriately). In this scenario, with *k*=21, a heterozygosity rate of only 1.19% is sufficient for the 1n peak to be as tall as the 2n peak.

Three genome profiling approaches also include estimates of genome-wide heterozygosity. First, in the genomes with low overall heterozygosity (<<0.5%), heterozygous sites will generally be more than *k* nucleotides away from any other heterozygous site. Consequently each heterozygous site will generate 2**k* heterozygous *k*-mers, which has been proposed as a straightforward estimate of heterozygosity [42]. However, many outbreeding species, hybrids, and other species display substantially higher levels of heterozygosity [45] [46]. The problem of linked variants (i.e. multiple heterozygous variants less than *k* bp apart) is addressed in GenomeScope. With the assumption that heterozygous loci as well as duplications are independent and uniformly distributed across the genome the fractions of heterozygous and homozygous *k*-mers are calculated as multiplication of per-nucleotide heterozygosity estimate (See supplementary materials of [11] for a detailed explanation with illustrations). The third approach, Tetmer, estimates the genetic diversity of a population and is based on an infinite site model and coalescent theory [[43,47,48]. For a more detailed comparison of the three methods, see **Supplementary Text 2**.

The most popular tool, and the tool we will use for all the examples below, is GenomeScope 2.0, which also includes support for polyploid genomes and more advanced model fitting methods [12]. We will use it to demonstrate various uses and problems with fitting genome models, but we encourage readers to also consider other genome-fitting tools, as those might be more suitable for their specific problems (See **Supplementary Table 2** for an overview).

## Common signatures of *k*-mer spectra

To generate a high quality reference of a diploid genome it is recommended to sequence at least 25-30x coverage of long reads, or more generally 15x per haplotype [49,50]. Even a simple visual inspection of *k*-mer spectra is valuable to quickly assess if this coverage is achieved. Such coverage should generate a *k*-mer spectrum that shows distinct coverage peaks as demonstrated by the European mistletoe *Viscum album* (**Figure 3A**). Sequencing data without sufficient coverage will have poorly defined peaks, because the homozygous and heterozygous genomic peaks will be blended at the left side of the coverage plot. If the peaks are still visible, it might be possible to fit a meaningful genome model, like in the case of the crayfish *Procambarus virginalis* (**Figure 3B;** data from [51]).

Usually, something already known about the species we sequence; in particular ploidy or genome size that were previously assessed via cytogenetic techniques. Confronting prior knowledge with the estimates derived from the *k*-mer spectra is often helpful in identifying potential problems in the data. In the case of the crayfish, there is a nearly perfect match of genome size estimate from the *k*-mer spectra and flow cytometry [51], supporting that the model converged well. Very low sequencing coverage or elevated rates of errors, leads to blending of peaks; genomic *k*-mers become indistinguishable from error *k*-mers. This is visible in the chive *Allium schoenoprasum* (**Figure 3C**) where the model (black line) does not fit the data (blue histogram) well. In such cases, the estimated values are just artefacts of a poor convergence. The predicted genome size is much lower than what we would expect in *Allium* genus, where other species have genomes ranging from 8.4-13.4Gbp [52], and the coverage is much higher than what we would expect from a spectrum of this shape. Coverage problems are usually resolvable with additional sequencing, while high error rates may require a different sequencing technology and/or library preparation.

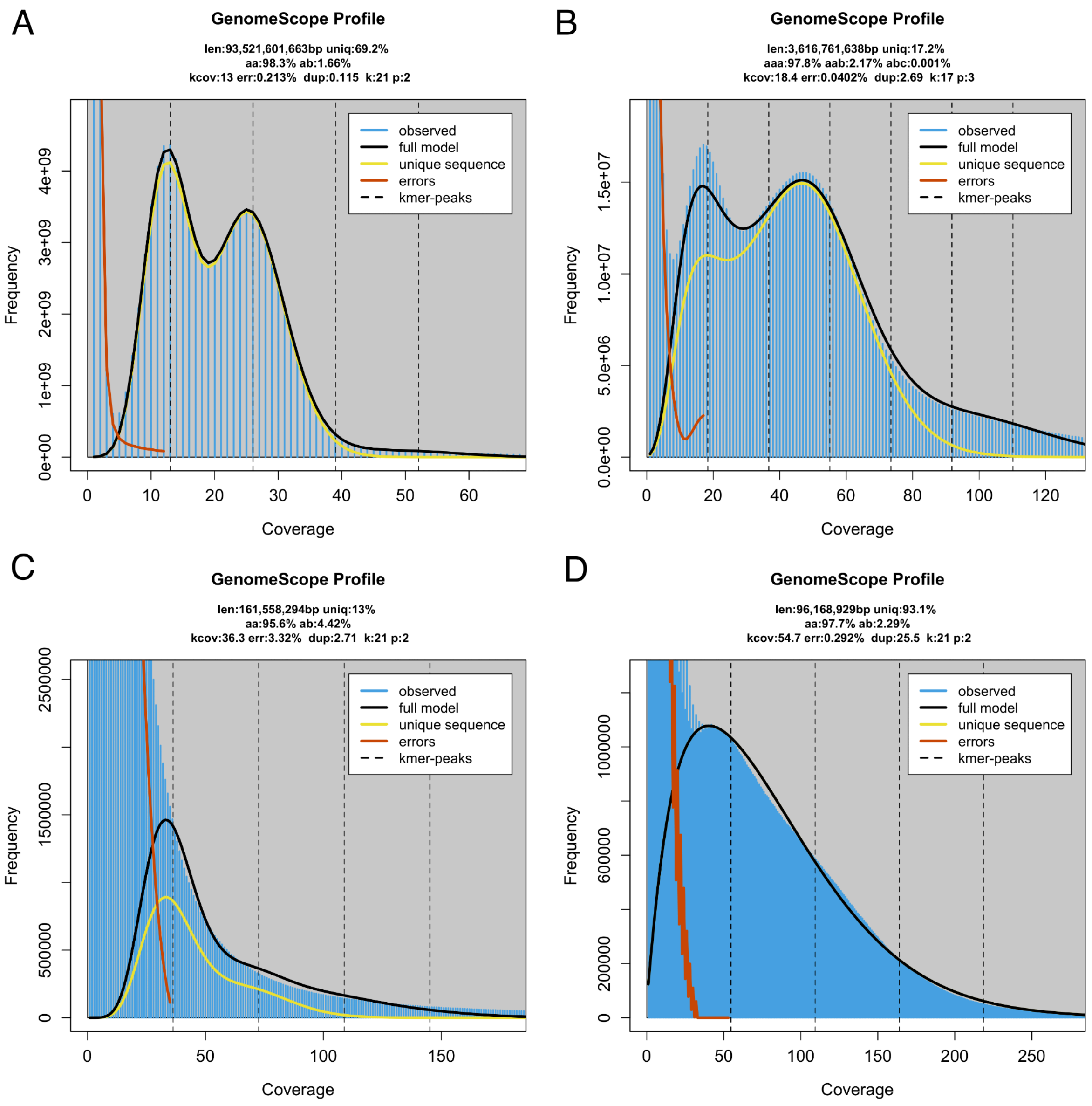

**Figure 3: Examples of *k*-mer spectra. A.** *Viscum album:* a diploid spectra with enough data to observe two distinct peaks and fit a model that accurately reflects genomic features despite the large size of the genome. **B.** *Procambarus virginalis:* *k*-mer spectra of a sample with low coverage, barely sufficient for a model fit. Notably, we used *k* = 17 to increase the *k*-mer coverage and make the model fit possible. **C.** *Allium schoenoprasum:* The sequencing coverage of this data set is approximately 1x, error *k*-mers and genome *k*-mers are completely blended, as a consequence the model did not converge to meaningful estimates. **D.** *Hypsibius dujardini* a heavily contaminated sample of a tardigrade.

Contamination is another common reason why we sometimes observe blended peaks in a *k*-mer spectrum, especially for small species that are difficult to culture in sterile environments or even impossible to culture at all [53]. In those cases additional species other than the target are sequenced along with the target sample, resulting in a contaminated genome profile. For example, the true genomic peak in the tardigrade *Hypsibius dujardini* is overlaid with several other co-biont genomes (**Figure 3D,** data from [54]). If a contamination consists of a very few species only and the coverage is very high, we might be able to see individual peaks corresponding to the individual sources. Those are, however, usually unevenly spaced (**Supplementary Figure 3**). Similarly to low-coverage datasets, the values reported by the genome model are most likely inaccurate.

Different sequencing datasets show different variation in sequencing depth, but the general rule is that higher coverage leads to better separation of peaks. Sometimes, we observe blending of peaks although the coverage is relatively high. This can be caused by various biological or technical reasons. For example, many bony fish sequencing runs show a *k*-mer spectra with a "bridge" blending the error *k*-mers and the 1n peak (**Supplementary Figure 4**). This pattern has been attributed to a high proportion of tandem repeats in many fish lineages [46,55] and a coverage dropout of low-complexity A/G rich regions of PacBio HiFi sequencing [56]. Another example is the well-known sequencing depth bias of Illumina short read sequencing suffers in regard to GC content [56,57]. This bias varies between chemistries [58] and is reduced with PCR-free strategies [59], but most prominently by the biology of the species.

The biases above are well described and replicable when using similar samples and sequencing technologies. However, sequencing runs are also affected by sample handling, used preservatives, or occasional manufacturing problems of sequencing flow cells. Even these problems are sometimes visible in *k*-mer spectra. For example, the *k*-mer based genome size estimate of a cape honeybee sample (142Mbp, **Supplementary Figure 5**) is much smaller than the published genome size (236Mbp). This was a case only for one of three samples, the other two samples from the same project (and same population) showed expected genome size. The spectra of the peculiar sample also show much higher blending of peaks, all together indicating this coverage dropout is not driven by biology of the sequencened specimen but rather technical difficulties.

In general, the lack of clearly defined and evenly spaced peaks indicates that there is no single source with sufficient coverage to generate the expected spectra. By keeping in mind the organismal biology of the target species many of the patterns and potential problems can be anticipated.

## Common pitfalls when fitting models

The quality of fit for genome models is largely dependent on the quality and coverage of the data, but also on the biological features of the genome. The most common problem of genome models is for the monoploid (1n) *k*-mer coverage to converge on a "wrong genomic peak". This usually happens if the 1n coverage peak is not sufficiently distinct or is significantly higher than the diploid peak. This can be caused by extremely low heterozygosity of the genome (i.e. the 1n signal is very weak, **Supplementary Figure 6A**) or because the coverage is very low and the 1n peak largely overlaps with the error peak (**Figure 3C**).

When the 1n coverage is not fitted correctly, none of the estimated values will carry any biological information regarding the genome. Therefore it is important to visually inspect fits and make sure the estimates agree with the context of the other known biology. For example, if we sequence a diploid selfing plant and the estimated heterozygosity is >5%, it is extremely likely that the true 1n coverage is ~½ of the estimated one. Selfing organisms are expected to have low genetic diversity which can confuse genome profiling tools to omit the first peak. In GenomeScope users can add the flag “-l <prior>” which allows the user to input a (1n) coverage prior and usually allows GenomeScope to converge to a biologically more relevant model (**Supplementary Figure 6B**). On the GenomeScope web interface (https://genomescope.org) this is labelled as “Average *k*-mer coverage for polyploid genome”. A typical process is to first run GenomeScope using default values without setting this parameter, inspect the plots for any potential issues, and then input a value for a second run if the original model fit is poor or does not match the expected biology for the species. Note the input value does not need to be a precise estimate, as the model fitting uses this to guide the automatic model fitting algorithm.

Another common issue is that many *k*-mer counting tools stop counting *k*-mer coverages at a user-specified value, with a typical default of 10,000. The last value in a *k*-mer histogram typically represents how many *k*-mers have the highest counted coverage, or higher. This will lose resolution of the frequency of the most common repeats in the genome. For many genomes it makes little difference, however, some extremely repetitive genomes will have a severely underestimated genome size. When using the default value, the estimated genome size of marbled crayfish is nearly half of the real genome size **(Supplementary Figure 7).** When one observes unexpectedly low genome size, it is frequently caused by the truncated *k*-mer spectra. Notably, FastK does not explicitly calculate coverage of every high-coverage *k*-mer, instead it reports as the last value the theoretical number of *k*-mers that would span the same total sum of coverage yield as all higher coverage *k*-mers in the genome, ensuring a more accurate genome size estimate [27,60].

## Joint interpretation of the *k*-mer spectra and the assembled genome

Post-assembly, estimated genome qualities from *k*-mer models can be used to assess assembly accuracy. One simple yet informative metric is to compare the estimated genome size to the assembly size. This simple comparison is quite powerful because it can quickly indicate misassembly, especially partial duplications caused by heterozygosity. *Ilex aquifolium* (common holly), sequenced by The Darwin Tree of Life Project [61] can serve as an example of mismatch between estimated genome size and assembly size. The estimated genome size of *I. aquifolium* based on *k*-mer histogram from PacBio reads is 815.6Mbp. However, the primary assembly is marginally larger (830.1 Mbp; **Supplementary figure 8A**) and furthermore there are 7.2% duplicated BUSCOs in this genome, altogether indicating that there could be some uncollapsed regions in this assembly with sequences from both haplotypes inflating the gene count and (haploid) genome size. This can be fixed downstream by purging duplicates, in this case it was done using purge_dups [61,62].

Several tools exist now to evaluate genome assembly accuracy using pre-assembly *k*-mers, we will specifically discuss Merqury, however this is not the only tool [37,63,64]. Essentially, these tools compare the *k*-mers from sequencing reads to the *k*-mers present in the finished genome assembly to determine how complete the assembly is. This operates under the assumption that the finished assembly should contain most of the *k*-mers that were present in the sequencing reads, excluding low coverage *k*-mers that are likely due to sequencing errors, and half of the heterozygous alleles in diploid genomes if we are working with a diploid collapsed assembly (see **Supplementary figure 8B** for Merqury plot example on the common holly genome). Additionally, Merqury estimates the consensus quality value score, which assumes all *k*-mers present in the finished assembly should be present in the read set. It then estimates the probability that a particular base is an error, which can be reported as the commonly used Phred score [65]. This approach is especially useful for species with no reference or a poorly assembled reference. However, it does require high-accuracy reads (such as Illumina or PacBio HiFi) as a reference point, and ideally these reference reads would be orthogonal to those used for the assembly.

## Comparison of sequencing libraries using *k*-mers

*K*-mers can also be used to identify large genomic differences between two sequencing libraries. This analytical approach can aid in answering a variety of research questions, most notably identifying sex chromosomes using libraries generated from different sexes. However, it can also be used to identify tissue or individual specific chromosomes such as B chromosomes [66]. *K*-mer chromosome identification techniques can be used on their own, or with well known techniques already in use such as methods that identify chromosomes based on coverage differences, or differences in sequence composition (e.g. SNP differences) between two sequencing libraries (reviewed in [67]). One advantage of *k*-mer techniques over other methods is a reduced reliance on a high-quality reference genome. Therefore, *k*-mer based techniques may be a better approach in non-model organisms with a fragmented or non-existent reference genome.

The basic approach in analyses using *k*-mers to identify specific chromosomes is to compare *k*-mer frequency in two sequencing libraries which differ in chromosome constitution. Generating a 2D *k*-mer spectra comparing *k*-mer frequency in two libraries allows for the identification of *k*-mers belonging to the chromosome with a different frequency in the two samples. For instance, if identifying a sex chromosome in a species with an XO sex determination system, *k*-mers belonging to the X chromosome will be at half the frequency in male compared to female sequencing libraries, while autosomal *k*-mers will be at the same frequency. The *k*-mers belonging to the X chromosome can then be isolated and either mapped to an assembly to identify scaffolds belonging to this chromosome or mapped to reads to identify reads belonging to the X chromosome.

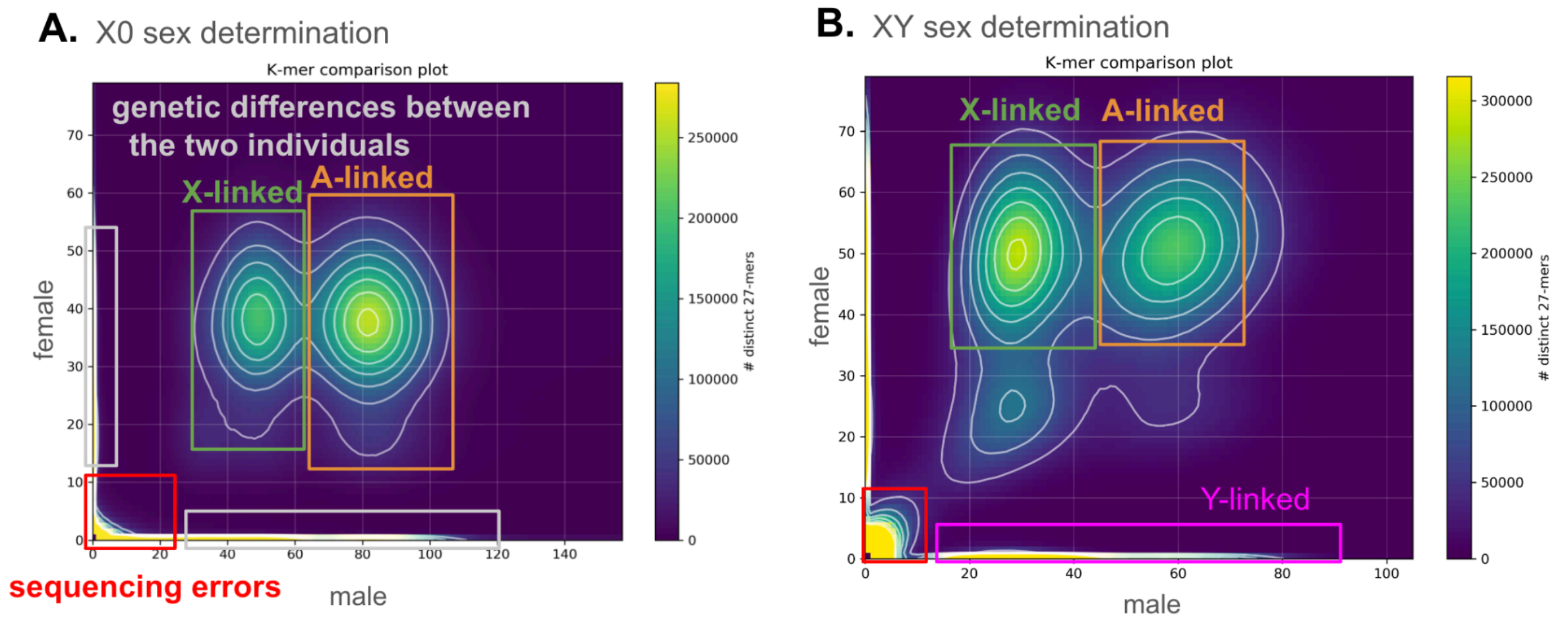

**Figure 4: *K*-mer comparison of multiple sequencing libraries.** An example of two 2D *k*-mer spectras for species with an **A.** XO sex determination system, and a **B.** XY sex determination system. Both plots show a heatmap of the frequency of *k*-mers in a male (x-axis) vs. a female (y-axis). Autosomal vs. X-chromosome *k*-mers in both plots can be differentiated as X-chromosome *k*-mers (green box) are at half the frequency in a male vs. a female sample compared to autosomal *k*-mers (orange box). Similarly, Y-chromosome *k*-mers can be identified in plot B as those that are absent in females but present in males (pink box). However, it is important to note that *k*-mers covering heterozygous sites on the autosomes of the male may be mistaken for those belonging to sex chromosomes.

Using *k*-mers to identify specific chromosomes is most effective when the region of interest is highly divergent from the other genomic regions. For instance, this approach is difficult to implement to identify the Y chromosome in a homomorphic XY sex chromosome system with low differentiation between the X and Y chromosomes, as few *k*-mers will be restricted to the Y chromosome. Therefore, situations such as this might require more permissive identification thresholds. The heterozygosity of the samples is also an important consideration. For instance, if identifying a sex chromosome from a male and female sample collected from a natural population, *k*-mers covering heterozygous SNPs will show signatures of *k*-mers specific to sex chromosomes as these will also be present at half the frequency of homozygous autosomal *k*-mers. This issue can be somewhat overcome by sequencing more individuals so that individual-specific heterozygous *k*-mers can be distinguished from sex-specific *k*-mers. Finally, an important consideration is the depth of sequencing for each sample. The sequencing coverage needs to be high enough to distinguish peaks in the *k*-mer spectra belonging to chromosomes at different ploidy levels in the samples sequenced and to distinguish from *k*-mers representing sequencing errors, similar to genome profiling.

# Some creative use cases of *k*-mers

Up until this point we have discussed general properties of *k*-mers and how to apply established *k*-mer methods to sequencing data. However, there are many problems in biology that are hard to solve with existing tools, which forces us to design novel approaches tailored to the problems and data we have. Here we showcase the creative use of *k*-mers in three examples of such problems in three different contexts: study of germline DNA, subgenomes of allotetraploid, and species assignment.

## Extraction of germline restricted chromosomes

A specific scenario of comparing libraries was used by Hodson *et al.* [67] to identify sequences belonging to the germline restricted chromosomes in the black-winged fungus gnat *Bradysia coprophila*, in order to explore the evolutionary origins of these unusual chromosomes. In this species, males show a peculiar set of elimination events that generate different karyotypes in different tissues and life stages (see [68] for recent review). Specifically, sperm carries two X chromosomes, two germline restricted chromosomes and one set of autosomes, while the soma has a single X chromosome, no germline restricted chromosomes (hence the name) and two sets of autosomes. In this case, germline tissue composed mostly of sperm has a different frequency of all chromosome types compared to somatic tissue. To characterise the germline restricted DNA of this species, approximately 95 testes of unmated males were pooled together in a single library, and sequenced separately from heads of the same individuals, to generate a clean somatic-tissue library to compare with the germline.

The comparison of the two libraries using 2D *k*-mer spectra indicated there was indeed an excess of *k*-mers that occurred solely in the testes library belonging to the germline-restricted chromosomes. However, unexpectedly the coverage of autosomal *k*-mers in the testes was still higher than the coverage of X chromosome *k*-mers, despite the fact that autosomes are at a lower frequency in sperm (expected 2:1 ratio X chromosome: autosomal *k*-mers in sperm). This was caused mostly by contamination of the germline library by somatic cells. A rough estimate indicated that ~77% of the germline sequencing library was composed of somatic cells, which caused the X chromosomes to have lower coverage than autosomes, but not ½, as you would expect if the sequencing library was just somatic tissue (**Supplementary figure 9**). Therefore, in this case, the fact that the two tissue types had three chromosome types all at different frequencies helped to determine the relative composition of each tissue type in the sequencing libraries.

The chromosome-group specific *k*-mers were matched to contigs, which resulted in near perfect assignment of contigs to chromosomal groups (germline restricted chromosomes, X chromosome, autosomes). In this analysis, the quality of the assignment was particularly good for two reasons. Firstly, because the germline restricted chromosomes happened to be very divergent from homologous regions on the X chromosome and autosomes (the somatic chromosomes), and secondly because this gnat line had extremely low heterozygosity as it was isolated by Charles W. Metz during the 1910s [68]. The *k*-mer based chromosome sorting allowed for accurate identification of tissue restricted chromosomes from a non-model species without a high quality reference genome (at the time), which facilitated downstream analyses of the origin of the germline restricted chromosomes [67].

## Separating subgenomes of allotetraploids

Polyploid genomes are challenging to assemble due to their large size, repetitive content that may arise from genomic shock, and the presence of chromosomes with similar composition (homeologs; or chromosomes that originated from speciation and were reunited in the same genome by allopolyploidization; [69] [69] [70]). Despite these challenges, several chromosome-level assemblies of allopolyploids have been published (e.g. [71] [72]), where subgenomes (i.e., the genomes that were reunited as a result of the polyploidization event) are reconstructed. In this context, *k*-mer approaches have been used to separate subgenomes, leading to significant advances in our understanding of genome evolution, including subgenome evolution, such as biases in gene retention, gene function, natural selection, and synteny [73,74] [72].

The separation of the two subgenomes is facilitated by the division of a lineages' evolutionary history into three tempos (**Supplementary Figure 10**): Tempo 1. The period preceding the speciation event that separated the two ancestral genomes; Tempo 2. The period between the speciation and the polyploidization events, during which ancestral genomes accumulated different transposable elements (TE) and diverged; Tempo 3. The period after the polyploidization event, during which both subgenomes coexist in the same nucleus. Now, consider TE accumulation in each of these separate tempos: Tempo 1. TEs accumulating in this tempo are expected to be evenly distributed on both subgenomes; Tempo 2. TEs accumulating after the speciation event and before polyploidization will be distinct on both subgenomes, as the two separate lineages will accumulate their own unique TEs; Tempo 3. TEs that accumulate after the polyploidization event will be roughly equally distributed between subgenomes ([74] [73] [72]).

In this context, *k*-mers serve as a powerful tool in subgenome separation in a chromosome-level assembly. Specifically, because evolutionary signals tend to overwrite each other on the genome, the fragmentation of the sequences offers an effective way to disentangle past processes, which have occurred in the three different tempos (**Supplementary Figure 10**) [72–74]. The separation of subgenomes involves two steps: first, the homeologs need to be identified. This can be done using UCEs/COS (ultra conserved elements, conserved ortholog sequence) or alignment-based approaches and synteny. When homeologs are known, *k*-mers allow identifying the accumulation of TEs in the three tempos. Specifically, by performing a *k*-mer spectrum analysis and searching for two signatures: first, selecting for *k*-mers which are present in high numbers (i.e. >100x) which should come from repeated areas on the genome, such as TEs; second, we select differently represented *k*-mers in members of the homeologs (i.e. *k*-mers which are found more than twice as often in one member compared to the other member). The assumption of high numbers targets TEs, whereas the second fishes out differentially represented TEs, thereby effectively focusing on the second tempo - when the ancestral genomes were separated and accumulating their own TEs. By doing a hierarchical clustering based on this *k*-mer separation, the chromosomes group on subgenomes.

## Species assignment using short *k*-mers

For specific types of analyses it can be beneficial to use very short *k*-mers ($k$<10). For the example discussed here [75], targeted amplicon sequencing was used to analyse haplotypes averaging only 160bp. The aim of the analysis is to identify the species by comparing the query sequences to a reference panel. As such, *k*-mers from the reconstructed haplotypes were analysed instead of the reads directly. Furthermore, because the haplotypes can be oriented by the primers, the analysis uses the full *k*-mer set rather than canonical *k*-mers.

The trade-off in the choice of *k* is between tolerance in sequence variation and captured complexity. Because the analysis works with reconstructed haplotypes rather than reads, the *k*-mer coverage does not play a role in the trade-off. For large *k* there is little tolerance for variation between the query and the reference, while for small *k* there is a high chance that the same *k*-mer is found in multiple locations in the sequence by chance. For example, in a 149 bp sequence, 5 evenly spread SNPs result in no 25-mers matching the reference. Conversely, the chance that all 4-mers are unique in a sequence of the same length is incredibly small ($<10^{-22}$). Based on these trade-offs, we selected 8-mers as a reasonable length. With a mean target length of 160 bp, the chance that all 8-mers within a haplotype are unique is 84%.

To perform species assignment, we compute the *k*-mer distance from the query haplotype to each haplotype in the reference panel. The *k*-mer distance quantifies the fraction of matching *k*-mers between query and reference . The Nearest Neighbour sequence is the reference haplotype that minimises the *k*-mer distance to the query haplotype. The species label is assigned by identifying the Nearest Neighbours for all amplicon targets of the query sample and aggregating their contributions to the assignment.

The amplicon panel and the species assignment method were developed to perform species assignment for the entire genus of *Anopheles* mosquitoes. *K*-mers provide an objective way to compare highly diverged sequences, where multiple sequence alignment or alignment to a single reference genome tends to introduce bias towards better-represented clades in the panel and the reference species respectively. Moreover, *k*-mers provide a natural way to incorporate small indels in addition to SNPs, which considerably increases the power to distinguish between species when working with less than 10kb sequence.

## The future of *k*-mers in genomics

As essential as words are to linguistics, *k*-mers are just as essential to bioinformatics. They are one of the most basic ways to represent a biological sequence, yet their utility cannot be overstated, as they are the fundamental data type used for a myriad of applications spanning genomics, transcriptomics, and metagenomics. Building on these successes, several research trends have emerged to make them even more efficient and effective for several important applications.

Within genomics, in addition to profiling increasingly complex biological samples, *k*-mers are now being used to power several genome assembly and analysis applications. One powerful technique has been the rise of “trio binning”, where *k*-mers identified in unassembled reads from parental genomes are used as markers to phase the unassembled reads of their children [76]. This enables genome-wide phasing and *de novo* assembly of the individual haplotypes in a sample, which has led to a renaissance in diploid genome assembly with improved contiguity and accuracy over prior approaches [77] . Another powerful technique has been a focus on “singly unique nucleotide *k*-mers” (or SUNKs) to aid in the assembly of repetitive genomes and repetitive sequences [78] . SUNKs can be identified within unassembled reads, and serve as unique sequences to “anchor” reads within a genome assembly with high confidence. They were essential, for example, to resolve segmental duplications in human genomes by identifying localised regions of unique sequence embedded within the complex repeat arrays [79]. Another clever application has been to use

*k*-mers as markers for variation to power genotype-to-phenotype association studies [80]. This is particularly powerful to tag and analyse structural variations as these are the most difficult class of variation to study from raw reads. Moving forward we anticipate future advances assembling and analysing more complex genomes and pangenomes using related *k*-mer based techniques.

Relatedly, *k*-mers also play a major role in a variety of sequence classification applications. Within metagenomics, the seminal algorithm kraken [81] exploits *k*-mers as signatures of individual species allowing fast and robust classification of individual reads without alignment. Within transcriptomics, the seminal algorithm sailfish [82] demonstrated accurate transcript quantification was possible without alignment by using *k*-mers as markers for individual transcripts. We expect many future applications for *k*-mers as markers for classification and quantification of diverse samples. We also anticipate future applications where *k*-mers are embedded into abstract semantic representations for machine learning applications, analogous to how word2vec [83] and related approaches have emerged as cornerstones in natural language processing.

Finally, one of the most important technical developments has been the rise of sampling and sketching techniques, such as minimizers [84] and minhash [85], to reduce the computational complexity of *k*-mers. The core idea of these approaches is that, for many applications, it is not necessary to exhaustively consider every possible *k*-mer in a sequence. Instead it is often sufficient to focus on a small representative subset for an analysis, typically representing a few percent or less of all possible *k*-mers. Because the subset is much smaller than the full list, it is substantially faster to compare the subsets and it requires much less memory. For example, minimizers were originally developed to reduce the computing requirements for genome assembly, and are now used to accelerate popular aligners like minimap2 [9] by focusing the algorithm to a smaller list of potential seeds. Relatedly, minhash powers the seminal mash [28] algorithm to quickly estimate the similarity between pairs of genomes by comparing small lists of representative *k*-mers, enabling thousands of genomes to be compared to each other on a laptop in a few minutes. The current research frontier for this work focuses on developing even more advanced schemes for selecting or combining *k*-mers together to ensure the representative subset of *k*-mers is unbiased while being robust to sequencing errors and repetitive elements ([86] for a review). There is also ongoing work to develop highly efficient software libraries and toolkits for the fast and efficient processing of *k*-mers for all these applications, especially to reduce the memory requirements for indexing large sets of *k*-mers [87] and indexing *k*-mers across large sets of genomes [88].

Overall, the future is very bright for continued research and application of *k*-mers in all forms. Whenever a researcher is considering profiling a genome, comparing genomes, or another alignment-based analysis, they should ask themselves if it might be accomplished using *k*-mers alone.

## Acknowledgements

This manuscript drew inspiration from the OH-KNOW workshop; we would like to thank Hugo de Boer and Quentin Mauvisseau from ForBio and other lecturers of the course - Siavash Mirarab, Rishi De-Kayne, Paolo Ribeca and Hannes Becher. We would also like to thank

Giulio Formenti, Konrad Lohse, T. Rhyker Ranallo-Benavidez, Arthur Delcher, and Adam Phillippy for their helpful discussions and feedback on the manuscript. This work is supported, in part, by National Science Foundation awards IOS-1758800 and IOS-2216612 (to M.C.S), Human Frontiers Scientific Program award RGP0025/2021 (to M.C.S), and KSJ was supported by the Wellcome Trust grant number 220540.

# Supplementary Materials: Guide to *k*-mer approaches for genomics across the tree of life

Note: References for supplements have their own reference list by the end of the document.

## Table of content

# Supplementary Figures

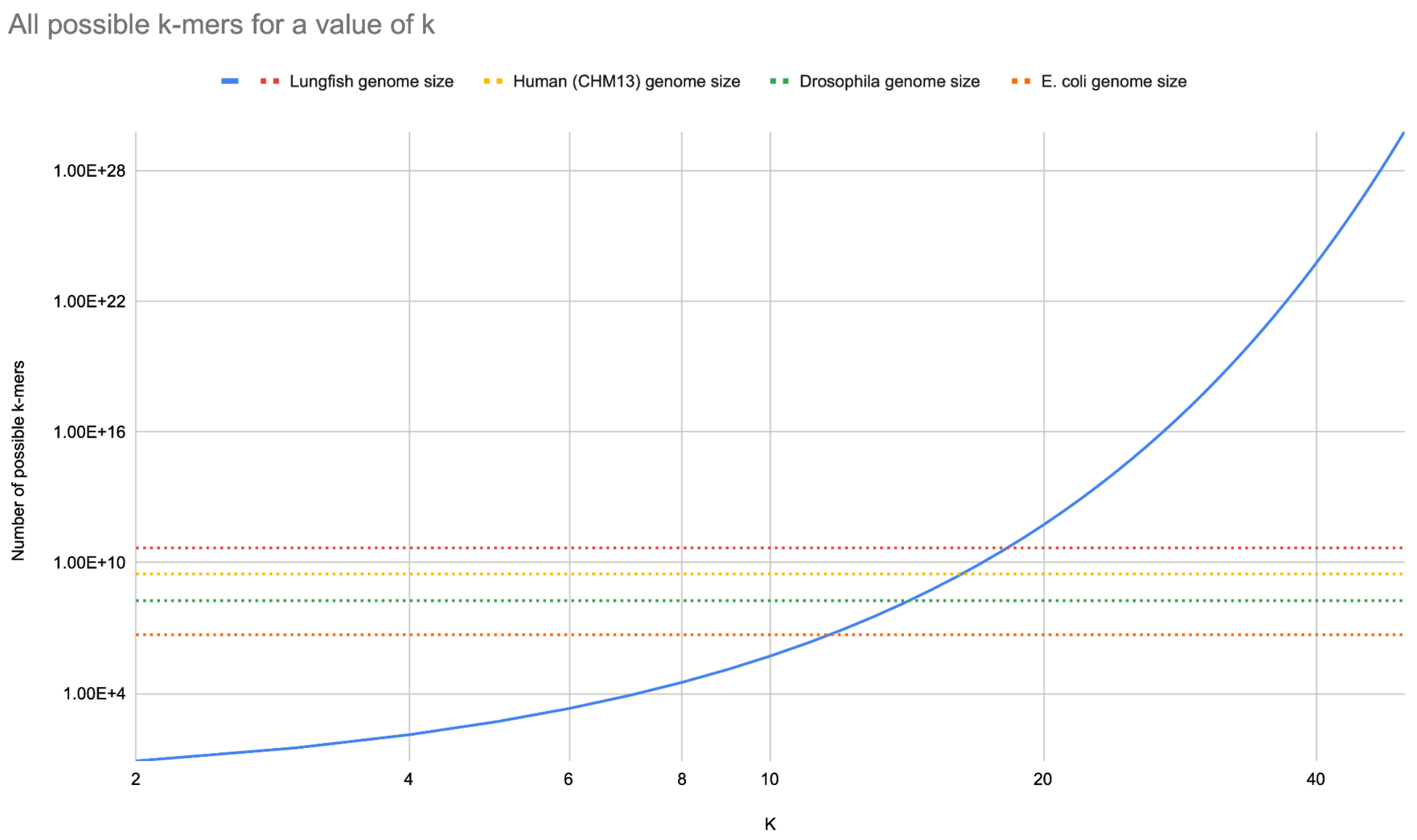

**Supplementary Figure 1: The total theoretical number of *k*-mers for each *k*.**

The appreciation of the exponential increase of the complexity of the *k*-mer can be illustrated relative to genome sizes of species. The proportion of *k*-mers corresponding to unique positions in the genome will increase with *k*. Specifically, there is a guarantee at least some of the *k*-mers are non-unique if the complexity of the *k*-mer space (blue line) is below the genome size (for $k < 11$). However, with $k > 21$, the vast majority of the regions that will share the same *k*-mer sequence are most likely to be similar for biological reasons (e.g. genomic duplicates).

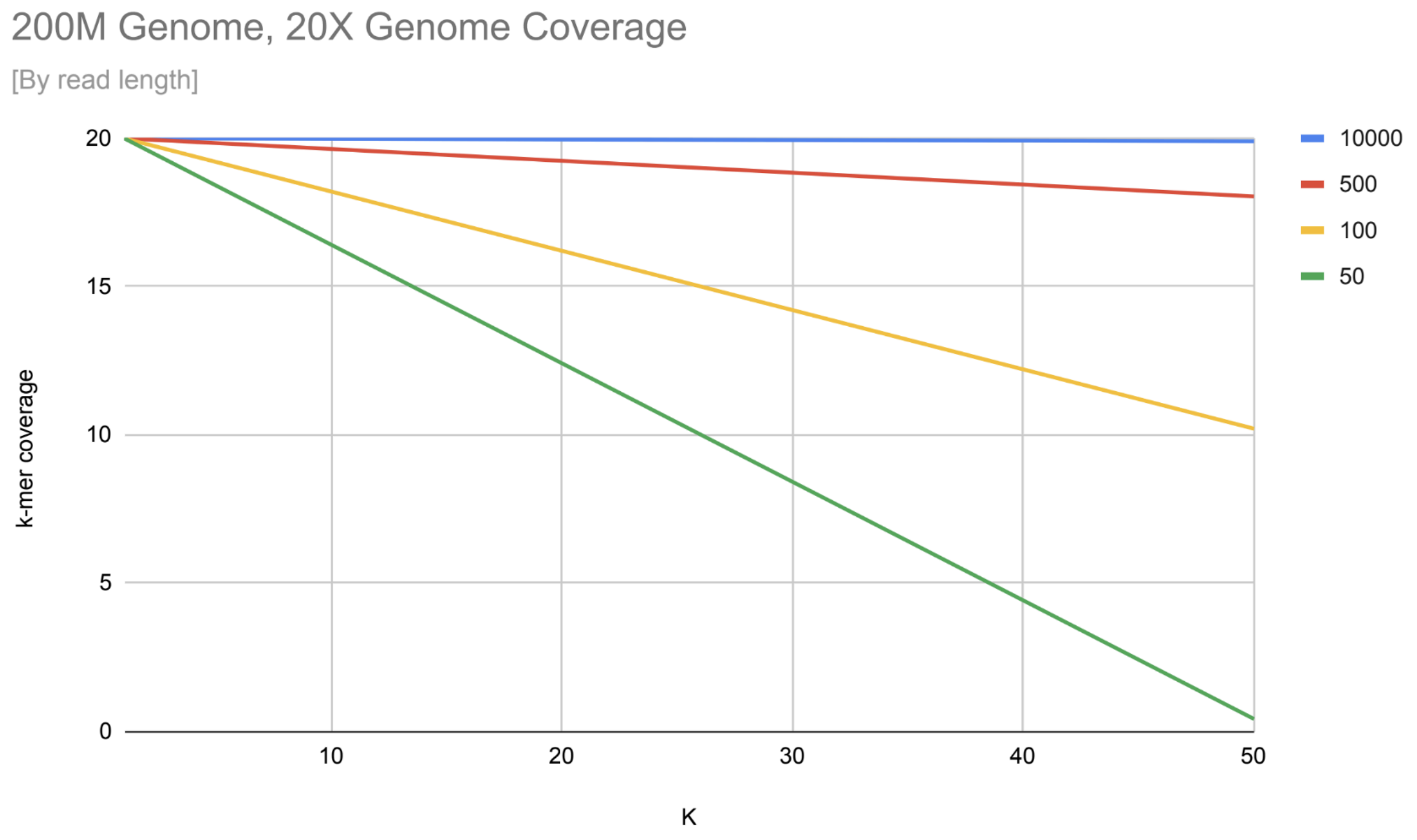

**Supplementary Figure 2: The relation of the *k*-mer coverage and *k* in relation to the read length**

This plot demonstrates the *k*-mer coverage that will be observed for different values of *k* for different length reads. Each line represents a different read length, and in each case 20x average genome coverage is available. When *k* is low, the difference is not very prominent; for longer *k* values, the greater the benefit of long reads to maintain high *k*-mer coverage. Note that this plot does not consider sequencing errors.

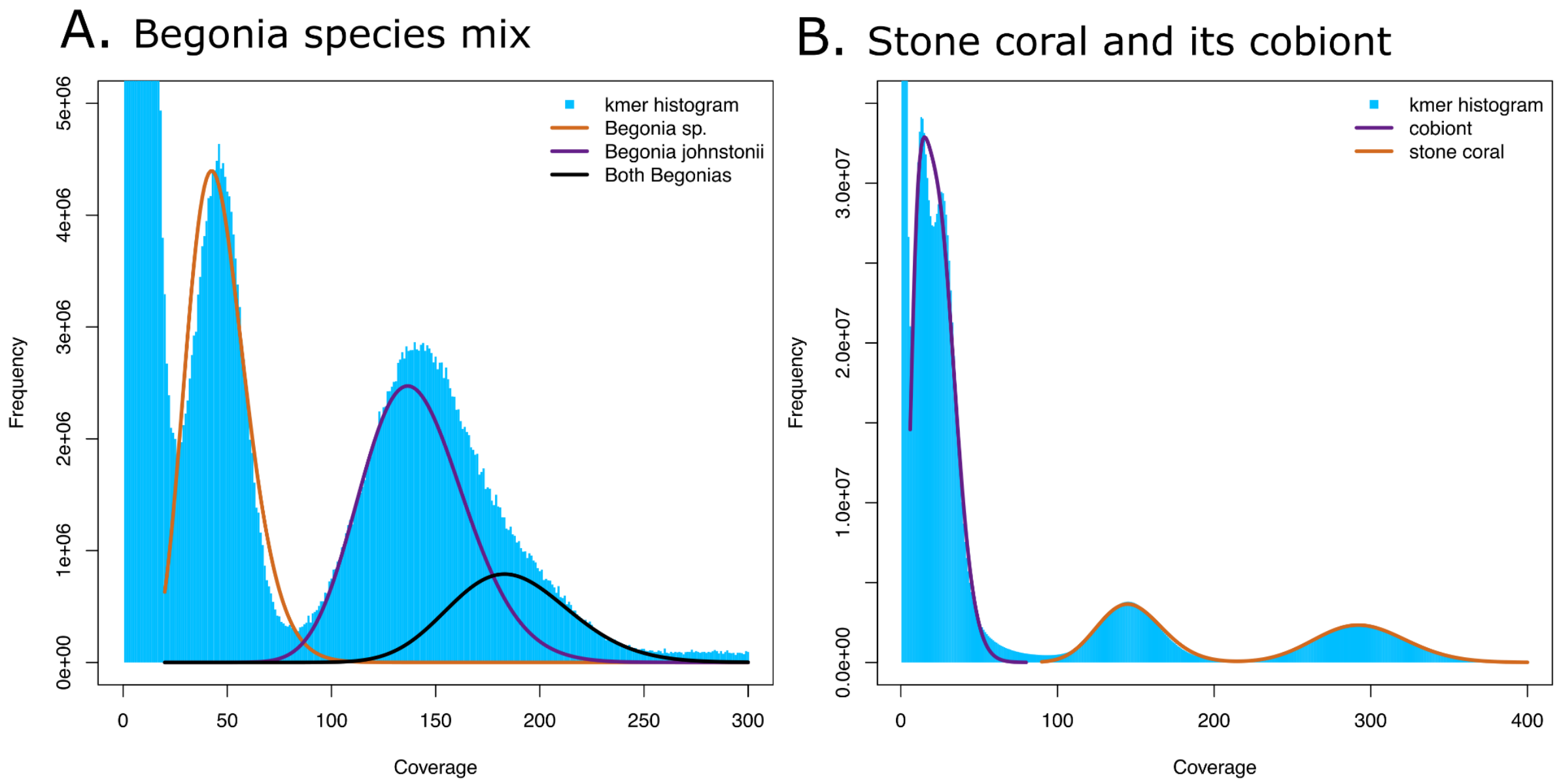

**Supplementary Figure 3: Example(s) of low complexity contamination visible in a *k*-mer spectrum.**

**(A)** This spectrum visually shows two peaks, but they are clearly not spaced in clear stoichiometry (1:2, 1:3 or 1:4). Instead, this spectrum is the result of mixing the genomic DNA of two different plants in the same sample. The central peak (purple) represents the *k*-mers unique to the sequencing target *Begonia johnstonii*, and the lower-coverage peak (orange) represents a different begonia species that grew in the same flower pot. The black distribution represents *k*-mers shared between the two *Begonia* species. This signature is typical for sequencing genomes of two closely related species. **(B)** This spectra shows a sequencing of a mixture of two different diploid genomes. The high coverage (orange) is the target genome of stone coral *Pocillopora grandis*, while the coverage peaks (purple) represent a cobiont. Data from: https://tolqc.cog.sanger.ac.uk/asg/jellyfish/Pocillopora_grandis/

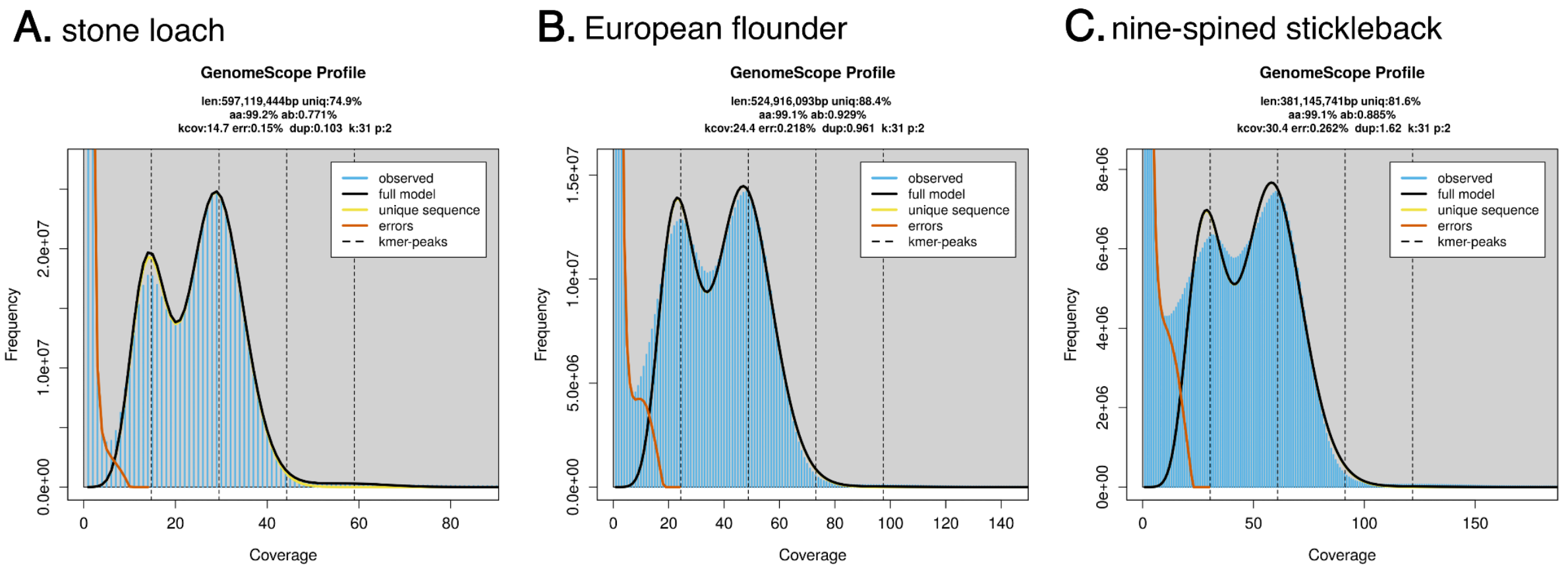

**Supplementary Figure 4: *k*-mer spectra of bony fish computed from PacBio HiFi reads**

All the spectra suffer from a dropout of sequencing coverage in some regions causing blending of coverage peaks despite a relatively high coverage. While the effect is very small in the stone loach [1] (**A**), it is more pronounced in the European flounder (**B**) and very apparent in the nine-spined stickleback [2]. This pattern has been associated with a reported coverage dropout of GA-rich low complexity regions in PacBio HiFi sequencing [3]. All three plots were retrieved from Tree of Life ToLQC portal https://tolqc.cog.sanger.ac.uk/.

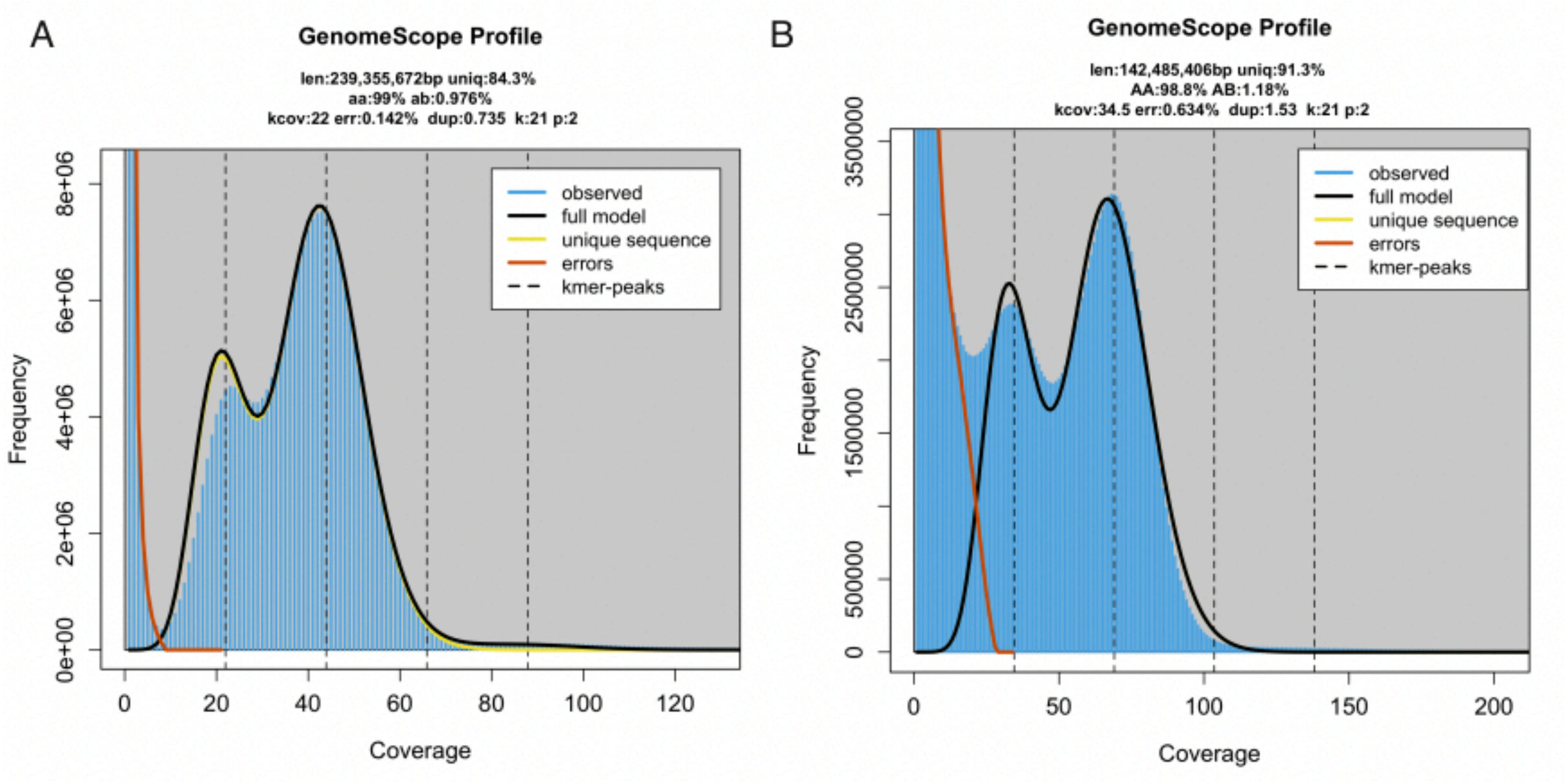

**Supplementary Figure 5: Genome proofing of two cape honey bee individuals.**

Data from [4]. **(A)** A *k*-mer spectra from a successful sequencing run: the error peak and genomic peaks are well separated and the predicted genome size is very close to the expected honeybee genome size. **(B)** Despite the greater *k*-mer coverage (~34x) the error peak and genomic peaks are not less separating indicating there was a problem with the sequencing library. This could potentially indicate contamination, but the genome size indicates instead there was a different problem as a large portion of the genome is missing. It is difficult to judge what exactly was wrong, but this sample is certainly not a high quality representation of a bee genome.

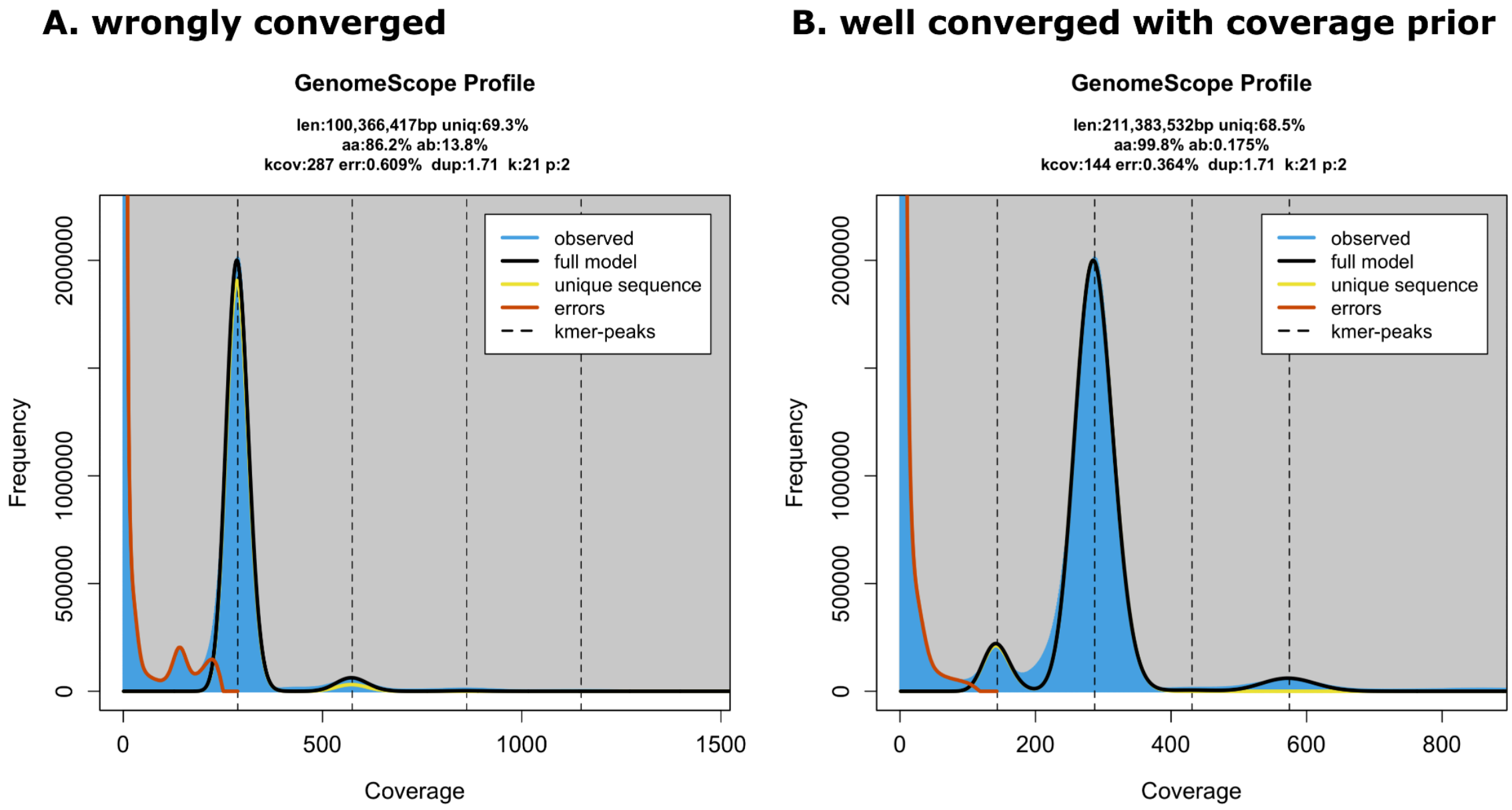

**Supplementary Figure 6: Genome convergence pitfalls**

The strawberry *Fragaria iinumae* is a diploid strawberry with a typical strawberry genome size around 200Mbp. This particular specimen (DRR013884) has low heterozygosity, which is also not unexpected in a strawberry. Data from [5] **A.** The default GenomeScope 2.0 run with no additional parameters misses the 1n coverage peak. One should spot several red flags: too small genome size, very high heterozygosity for a strawberry and finally, the red error line shows what could be (and really is) another genomic peak that is not part of the model **B.** GenomeScope 2.0 run with specified coverage prior (-l 140) converged on correct peaks generating biologically meaningful estimates of genome properties.

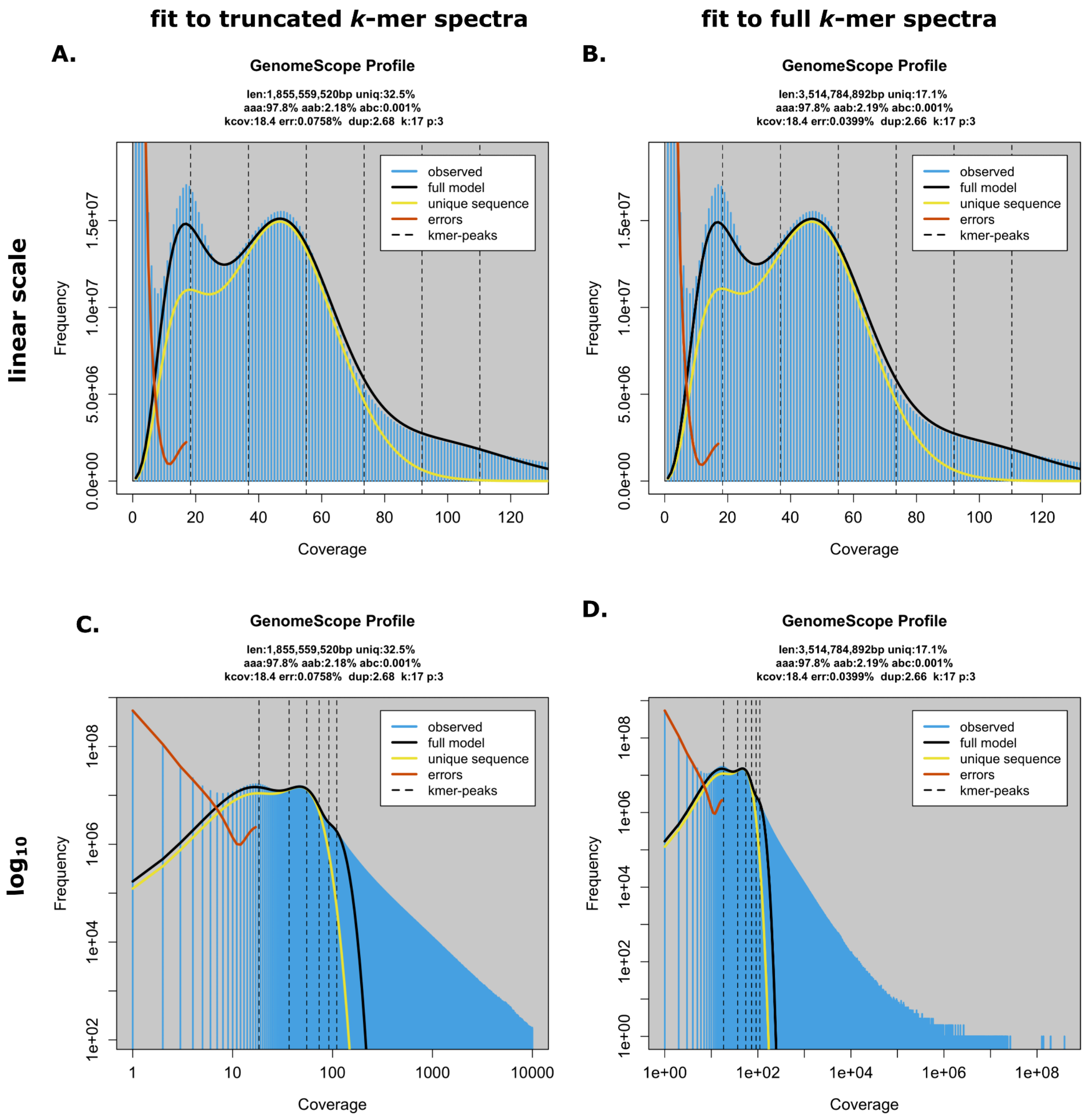

**Supplementary Figure 7: Genome size estimate pitfall**

The Marbled crayfish has a triploid genome with ~3.5Gbp genome size measured by flow cytometry [6]. Data from [6]. **A.** When using the default threshold by KMC, the *k*-mer spectra is truncated at 10,000 coverage. The haploid genome size estimate is then 1.855 Gbp, which is approximately half of expected haploid size **B.** When increasing the coverage threshold to 500,000,000, the estimated genome size (3.515 Gbp) is close to the expected values given the flow cytometry measurements, indicating that a substantial proportion of the genome is on extremely repetitive sequences, which can be observed on the $\log_{10}$ scale plots (**C.** and **D.**).

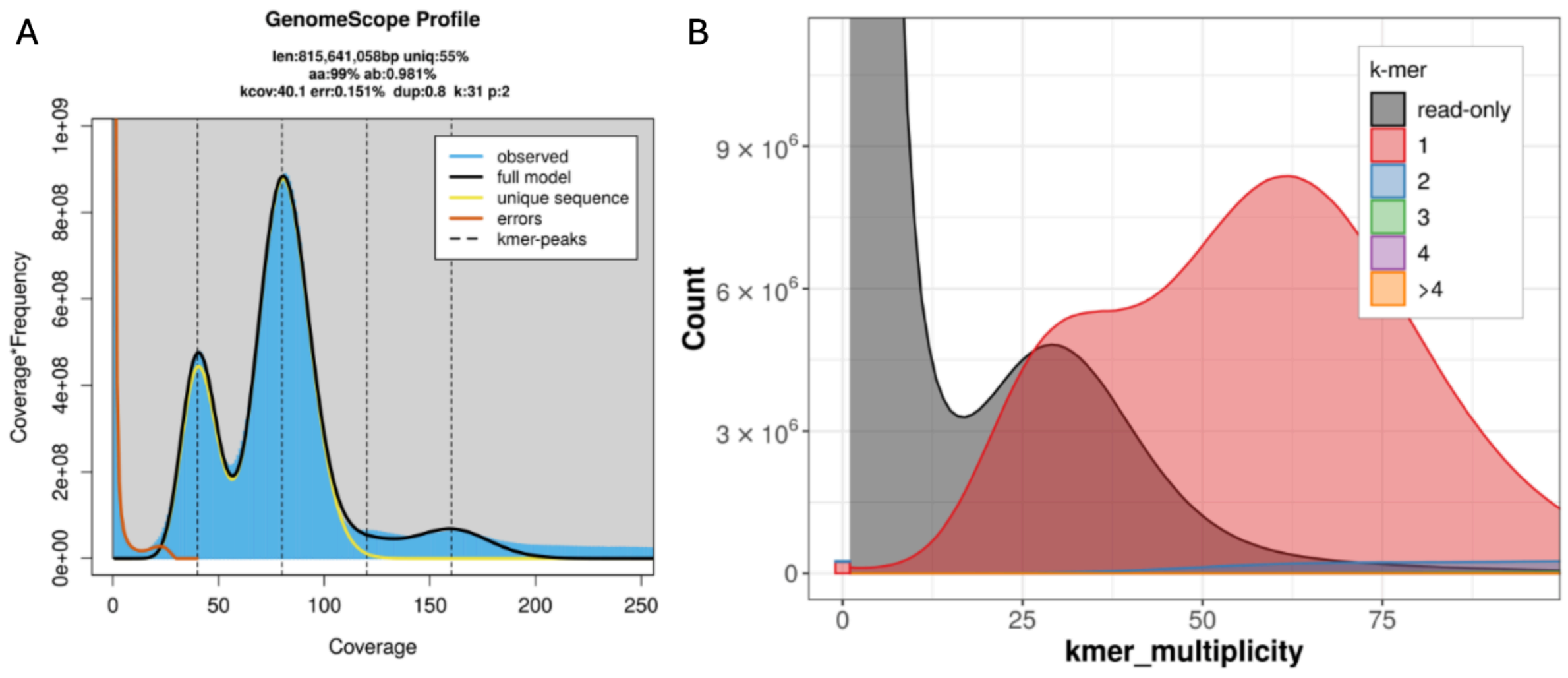

**Supplementary Figure 8: example(s) of assembly quality assessment using the *k*-mer spectrum.**

**(A)** The PacBio HiFi 31-mer spectrum of *Ilex aquifolium* from DToL. The model fit indicates the genome is around 815.6 Mbp, with a fairly high level of heterozygosity (~1%). **(B)** Merqury plot for assembly QC of the same species. The black area represents the *k*-mers present in the read set but not in the assembly, and the red area represents what is present in the assembly once. Larger assembly size than the genome size estimate, higher 1n peak in the assembly (red) than absent (black) and relatively high BUSCO duplication scores all indicate there are uncollapsed haplotypes in the haploid genome assembly which will required downstream haplotype collapsing. Data from https://tolqc.cog.sanger.ac.uk/darwin/dicots/Ilex_aquifolium.

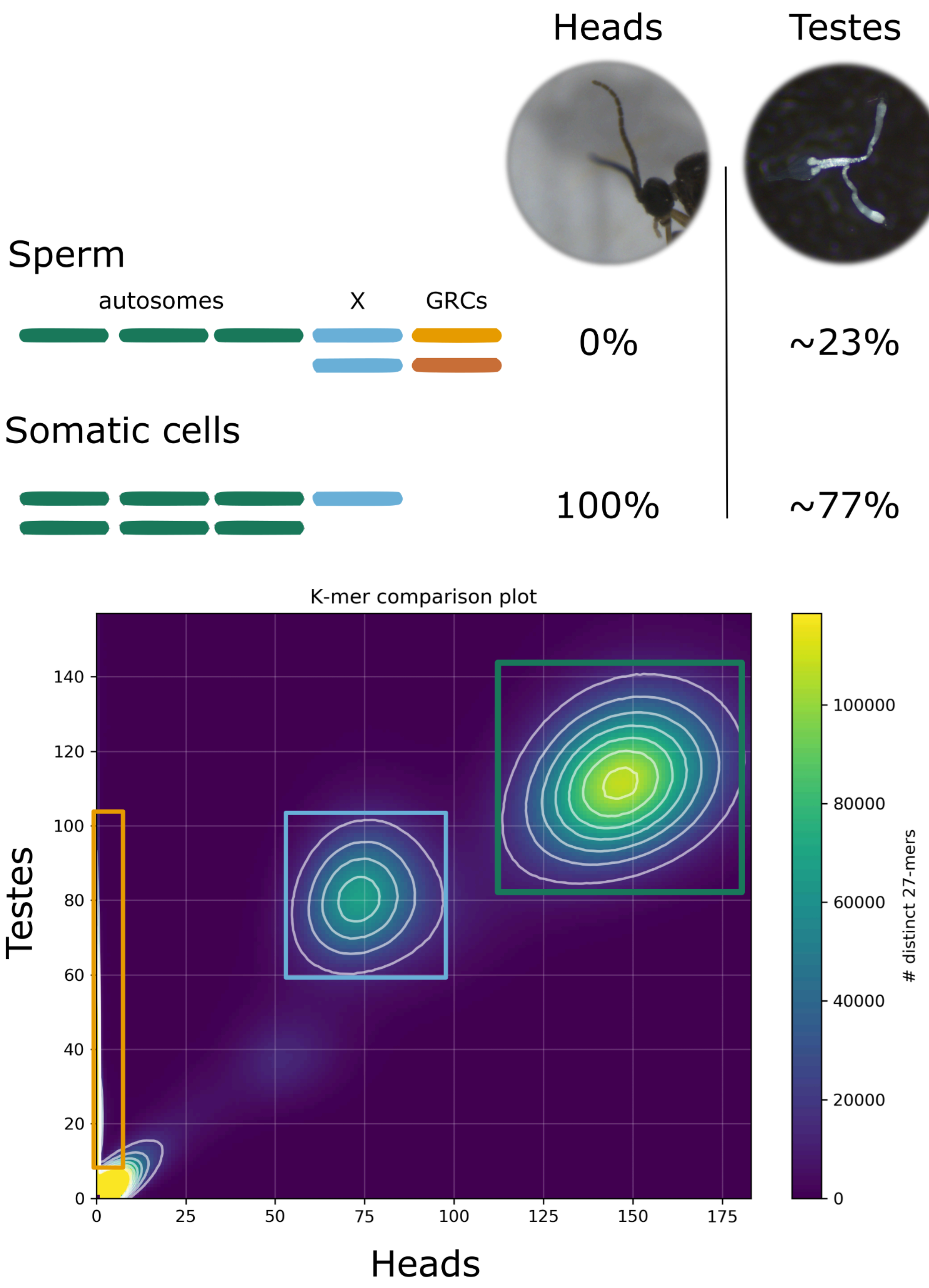

**Supplementary Figure 9: Comparison of head and testes libraries in species with germline restricted chromosomes**

The two compared libraries are heads and testes. While heads are pure somatic cells with uniform karyotype, the testes are a mixture of somatic cells surrounding the germ-line and germline consisting mostly of sperm. The approximate proportion of sperm and somatic cells were calculated using mean coverages of X chromosomes and autosomes. Sperm cells contain two diverged germline restricted chromosomes (GRCs), that show in the 2D *k*-mer plot in the orange square.

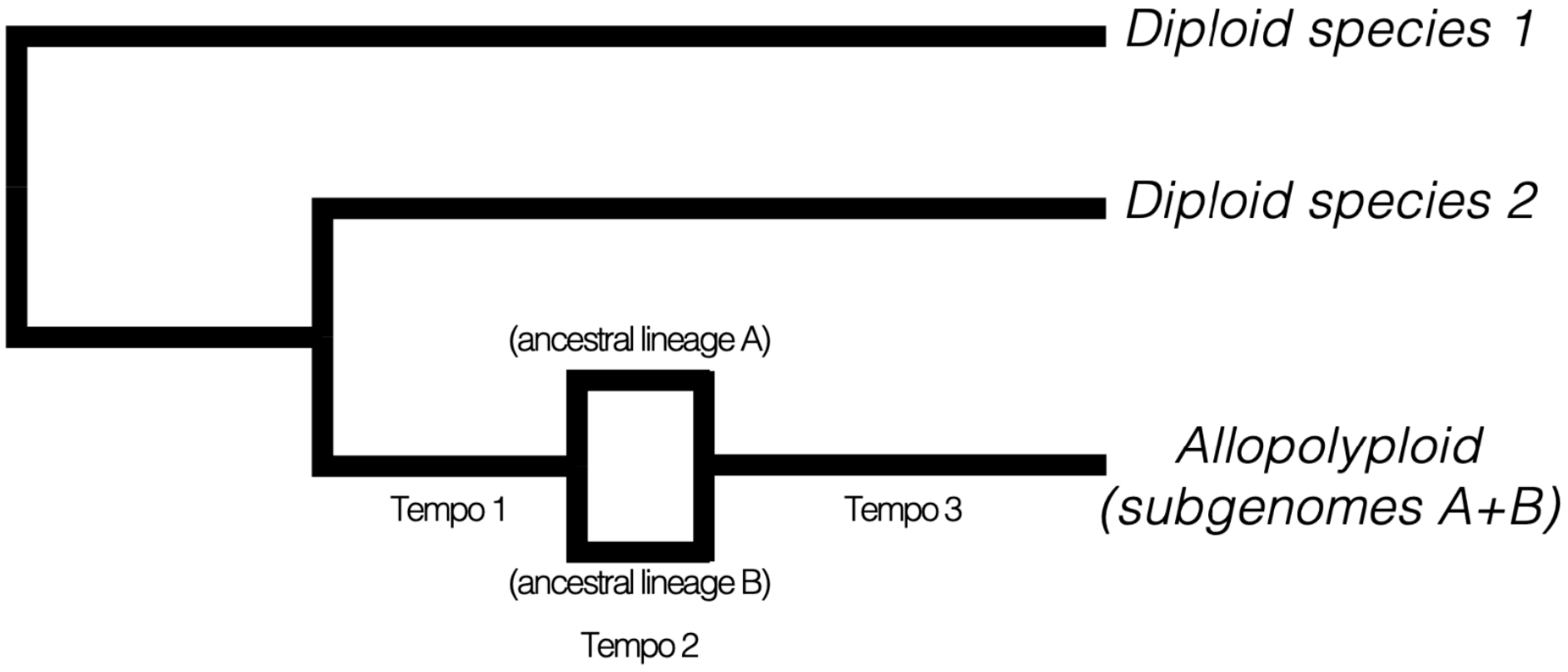

**Supplementary Figure 10: Phylogenetic representation of an allotetraploid species and the accumulation of transposable elements.**

Tempo 1 occurs after the speciation event, between diploid species 2 and the allopolyploid. The accumulation of transposable elements during tempo 1 will be evenly represented across both subgenomes. Tempo 2 represents a period where lineages of the allopolyploid are segregated and accumulate differences. Transposable elements accumulated in tempo 2 will be unique to each subgenome. The third tempo begins with the polyploidization event. Transposable elements in this period will be common to both subgenomes.

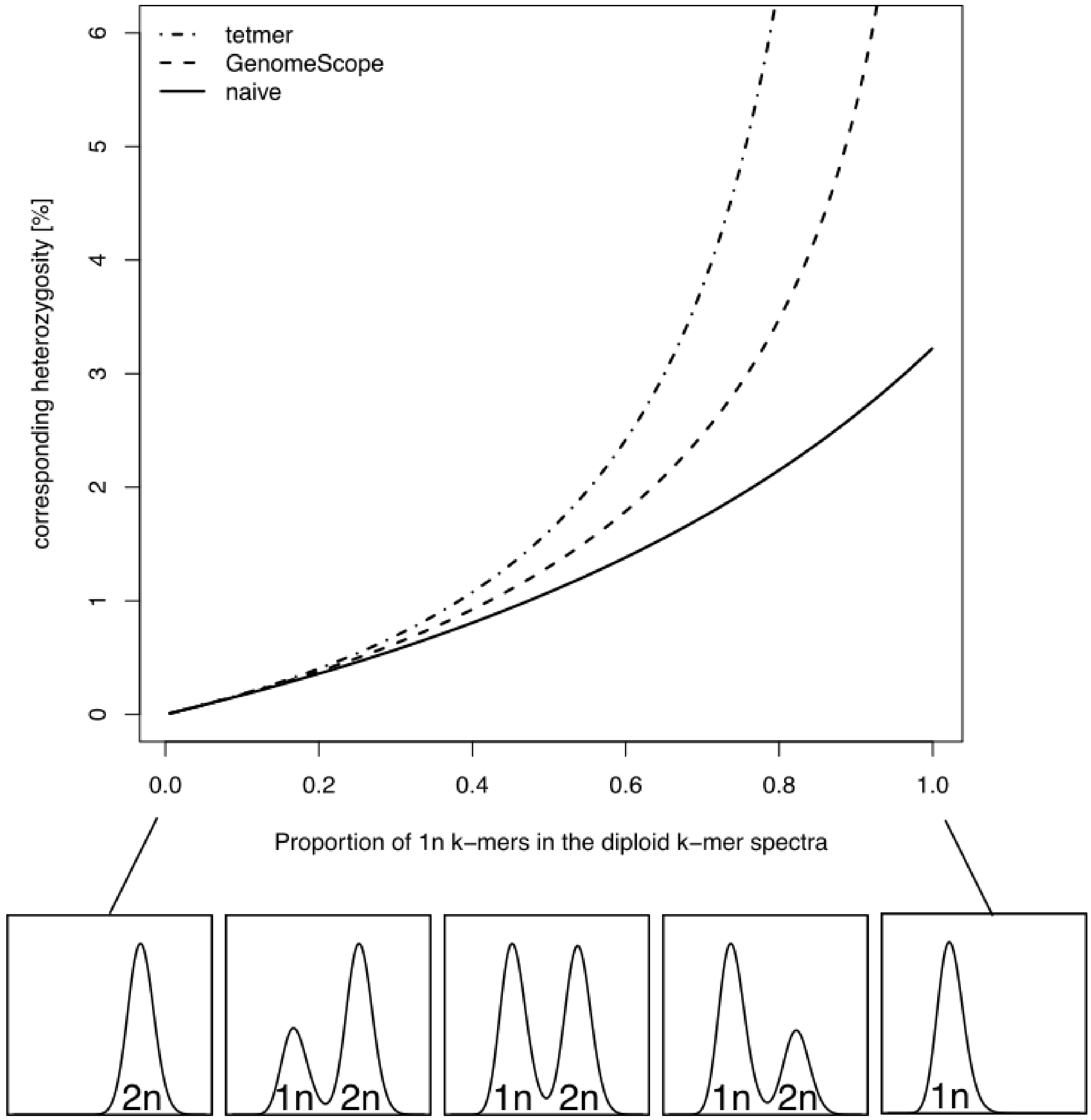

**Supplementary Figure 11: Sizes of peaks corresponding to heterozygosity estimates by different methods**

The three available methods estimating heterozygosity differ in the interpretation of the relative size of the 1n peak. In the case of a small 1n peak (<0.2 of all the *k*-mers), all methods estimate similar levels of heterozygosity. With an increasing proportion of *k*-mers in the 1n peak, the estimate starts to differ. Notably, Tetmer does not estimate heterozygosity, but genetic diversity, which is not exactly the same measure.

# Supplementary Tables

**Supplementary Table 1: Overview of selected *k*-mer counting tools and tool kits**

The is not an exhaustive list of *k*-mer counting tools, but an overview of those that are either historically or currently impactful. Most of the tools have their own specialised utilities no other tools does - all of them were developed with a clear intention (see the references).

| *Tool Year* | *description* | *url* | *reference* |
|---|---|---|---|
| *Jellyfish 2011* | *Very popular and easy to use* k*-mer counter, but substantially slower compared to more recent* k*-mer counters* | *https://github.com/gmarcais/Jellyfish* | *[7]* |
| *Khmer 2015* | *A versatile* k*-mer suite that allow* k*-mer indexing, streaming and building de Brujin graphs* | *https://github.com/dib-lab/khmer* | *[8]* |
| *KAT 2016* | k*-mer counter that is also able to compare two libraries or a library and a genome, powerful diagnostic plots of* k*-mer composition vs coverage* | *https://github.com/TGAC/KAT* | *[9]* |
| *ntCard 2017* | *Very fast, but approximate* k*-mer counter. The* k*-mer counting is just approximate and limited to non-repetitive portion of the genome* | *https://github.com/bcgsc/ntCard* | *[10]* |
| *KMC3 2017* | *Very fast explicit* k*-mer with the capacity to calculate any arbitrary coverages users request by parameters; Interface is a bit more complex* | *https://github.com/refresh-bio/KMC* | *[11]* |
| *Meryl 2020* | *A counter that also implements efficient operations between sets of* k*-mers, like union, intersection, subtraction. Used by Merqury.* | *https://github.com/marbl/meryl* | *[12]* |
| *FASTK 2023* | *The fastest explicit* k*-mer counter, compatible with all the existing downstream tools; using a shortcut to aggregate all repetitive* k*-mers (explicitly counting only to* $2^{16}$*x coverage).* | *https://github.com/thegenemyers/FASTK* | *[13,14]* |

**Supplementary Table 2: List of selected genome profiling tools**

| *Tool* | *description* | *url* | *reference* |
|---|---|---|---|
| *Kmerfreq* | *Estimates genome size from corrected long reads.* | *https://github.com/fanagislab/kmerfreq* | *[15] [16]* |
| *RESPECT* | *Primarily for skimming data; ignores the concept of heterozygosity and is pre-trained on haploid representation of genomes* | *https://github.com/shahab-sarmashghi/RESPECT* | *[17]* |
| *findGSE* | *Modelling* k-*mer coverage with the most complex distribution of all methods - skew normal distributionl,* | *https://github.com/schneebergerlab/findGSE* | *[16,18]* |
| *tetmer* | *focused on polyploids; based on coalescence theory* | *https://github.com/hannesbecher/shiny-k-mers* | *[19]* |
| *GenomeScope* | *using combinatorics and probability;* | *https://genomescope.org* | *[20]* |
| *GenomeScope 2.0* | *using combinatorics and probability; enhanced model accommodates polyploids; fit to transformed* k-*mer spectra* | *https://genomescope.org/genomescope2.0/* | *[13]* |
| *CovEst* | | *https://github.com/mhozza/covest* | *[21]* |

# Supplementary Texts

### Supplementary Text 1: The history of calling *k*-mers "*k*-mers"

The concept of decomposing sequences of letters in all possible subsequences is widespread across multiple disciplines - from linguistics, through information theory to genomics and biology. Being so useful, this concept was developed independently several times under many different names. Computational linguists called the substrings -grams, mathematicians called them -tuples or -tups, in biology, the most frequent expression is -mer. Some authors recognized the difficulty in communicating the concept, so they decided to use simply -words instead, but unfortunately that led to even more confusion. In many cases, the substrings had specific length, and then authors would use concrete numbers as prefixes, e.g. 11-mer for a polymer of length 11. But sometimes, the length was just a variable, so people used various letters to mark the unknown and these letters also vary a lot.

**-grams**

Probably the oldest reference to the concept dates all the way back to [22,23], Shannon used "N-grams" to develop a theory for communication, later to calculate entropy of a natural language. This is likely the oldest record of the *k*-mer concept (in the sense of all possible substrings of a certain length). The concept received a lot of appreciation in the 1990s for applications as approximate string matching, and quite often referred to as "q-gram". In 1999, a tool QUASAR was published - q-gram based database search using suffix arrays [22].

**-tuples**

In maths, tuples are ordered sets of elements. The individual letters are the elements, but it is their order that is really important for defining each subsequence - this definition emphasises that ATGA is not the same sequence as AATG although it has the same elements. A shorter version of this notation (ktup, controversially, without a dash between of k and tup) was used in the, these days legendary, FASTA method for amino-acids sequence alignment [24]. There was a lot of work done on -tuples till early 2000s when they were slowly replaced by *k*-mers. Interestingly, authors working on -tuples used all sorts of prefixes: k-tuple [25] [26] REF, L-tuple [25] or ℓ-tuple [27]. The transition from k- to L- happened in [25], where they use k-tuples are a theoretical string and L-tuples are all the possible sequences of the length of a genome (which would allow perfect sequencing by hybridization, for the record, this will be one of those numbers higher than the number of the atoms in the universe for even a modest genome).

**-words**

This expression was introduced by a research group led by Waterman in early 2000s [28], even specifically with the k- prefix [29,30]. Perhaps one curiosity a careful reader might have noticed is that the very same research group used various forms of -tuples in the past (see the section above), so perhaps the best explained as an attempt to make the concept more accessible.

**-mers**

Finally, the most common term in bioinformatics these days is *k*-mer, which is simply for polymer of length k, although it is hardly ever used for anything else than a nucleotide sequence. The first record of -mer I found was from 1968 by Mandeles [29]. With understanding that string of nucleotides might have a unique position in the genome - they called these oligonucleotides unique-mers. Namely, they were placing two uniqe-mers referred as Ψ-mer and Ω-mer respectively. Two decades later, sequencing by hybridization (SBH) was proposed as a new alternative to sequencing on gels; The idea was to hybridise the sequence on a chip with short nucleotide probes (5, 8, or 10 nucleotides); The only challenge was losing the positional information, which naturally created the problem of "*k*-mers" - unplaced genomic substrings.The 11 bases long nucleotide sequences were called 11-mers [31]. Which became the standard for the following papers on the topic. One notable exception is the original description of microarrays, where they were referred to as 15 nucleotide oligomers [32], however then also used "15-mer" in the product descriptions once the commercial product (Affymetrix chips) were released. These are sequences of a specific length, not conceptually utilising the idea of taking sub-sequences of any arbitrary length. Which is of course understandable, that is a pre-sequencing era. The true *k*-mers appeared in the publication of BLAST in 1990, however using w as a prefix (w for word), so "w-mers" [33]. Nor w-mer or *k*-mer received too much attention in this era. Majority of people using this concept were coming from mathematical or computer science backgrounds and used other terms mentioned above. The use of "*k*-mer" became more common in late 1990s, including within the seminal work of MUMmer in 1999 [34]. In 2000, Liu & Singh coined "*k*-mer word frequency distribution" and described it as a "signature" of the sequence [35]. One of the other pioneers of expression "*k*-mer" were Mullikin & Ning in the Phusion Assembler publication [36]. In the paper they use this expression as well as plot "word frequency graph", which is one of the earliest *k*-mer spectra plots [36]. Publications using the word *k*-mer increased in the following years compared to any other of the terms. This gradual process was likely completed with the release of several tools including a very popular *k*-mer counter Jellyfish ([36,37]. This counter served for a long time as the goto *k*-mer counter and likely played a role in solidifying the expression "*k*-mer" as the main way to talk about this concept.

## Supplementary Text 2: Corresponding fractions of heterozygous *k*-mers and heterozygous nucleotides

The main text outlined the principle that heterozygous sites in a genome generate twice as many *k*-mers of half coverage than homozygous sites. Specifically, if the variant is a SNP and no other variant occurs within *k* nucleotides, there will be 2 * *k*, heterozygous *k*-mers generated: *k* *k*-mers from the maternal allele, and *k* *k*-mers from the paternal allele. This is the **simplest model** on how to model heterozygosity, but is not realistic for moderately or highly heterozygous genomes.

The **GenomeScope** model overcomes this problem by considering the probability of observing homozygous and heterozygous *k*-mers respectively given a single per nucleotide heterozygosity parameter r - the probability of observing a heterozygous nucleotide [20]. In a simplified case (modelling two peaks only) the probability of observing a completely homozygous *k*-mer is $(1 - r)^k$, which can be complemented by the probability of observing a heterozygous *k*-mer pair $1 - (1 - r)^k$ (and given two *k*-mers are generated, there will be a factor 2 in the model fit to a *k*-mer spectrum) [20]. Important here to note, that first GenomeScope fits 4 peaks also including duplications, and considers random overlap of duplications and heterozygous sites (see supplementary materials of [20] for a very well-illustrated explanation). The GenomeScope 2.0 expanded this up to hexaploidy, and added many important features that further improved the fit, however, the fundamental logic of the fit remains the same [13]. The GenomeScope model is still somewhat “unrealistic” for several reasons: different regions within genome have different probability of being heterozygous (i.e. heterozygosity is not uniformly distributed in a genome); many variants are not just SNPs; and/or a large proportion of the genome might be covered by repetitions with more than two copies. How much of a problem this presents in the estimates is still an open question, and the answer is most likely dependent on the studied species.

Finally, **Tetmer** [19] is a tool that estimates genetic diversity (θ) as opposed to heterozygosity (fraction of nucleotides that differ between haplotypes). Specifically, the method assumes no *k*-mer recombines within, and the probability of a homozygous *k*-mer pair is $\theta_k / (\theta_k + 1)$, where $\theta_k$ is a per-*k*-mer genetic diversity. It is suggested for $\theta_k$ to be simply divided by *k* to obtain per nucleotide, assuming no overlap of variants. This method is more interesting for tetraploid cases with two different coalescent models for auto- and allo- tetraploid species respectively.

A different way to look at the difference of the three methods is “what heterozygosity would be predicted given the same relative size of the 1n *k*-mer peak”. In this comparison, we can easily see that the GenomeScope estimate will always generate higher heterozygosity estimates compared to the simple model, but smaller than Tetmer (**Supplementary Figure 11**). While looking at the plot, note that Tetmer does not use the same type of estimate, which makes the comparison somewhat unbalanced.

# Supplementary References